\documentclass{article}

\usepackage[utf8]{inputenc}
\usepackage[T1]{fontenc}

\PassOptionsToPackage{table, dvipsnames}{xcolor}
\usepackage{xcolor}

\PassOptionsToPackage{colorlinks=true, citecolor=blue, linkcolor=blue, urlcolor=black}{hyperref}

\usepackage{arxiv}
\usepackage[utf8]{inputenc}
\usepackage[T1]{fontenc}
\usepackage{url}
\usepackage{booktabs}
\usepackage{nicefrac}
\usepackage{microtype}
\usepackage{lipsum}
\usepackage{graphicx}
\usepackage{natbib}
\usepackage{doi}
\usepackage{algorithm}
\usepackage{algpseudocode}%
\usepackage{listings}%
\usepackage{multirow}
\usepackage{multicol}
\usepackage{amsmath,amssymb,amsfonts}
\usepackage{amsthm}
\usepackage{mathrsfs}
\usepackage{float}
\usepackage{bm}
\usepackage{caption}
\usepackage{hyperref}

\usepackage{tabularx}
\usepackage{array}
\usepackage{pdflscape}

\title{Generative wave propagator
}

\author{
	\href{https://orcid.org/0000-0001-8868-7967}{\includegraphics[scale=0.06]{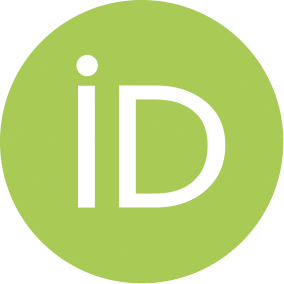}\hspace{1mm}Shijun~Cheng} \\
	Division of Physical Science and Engineering\\
	King Abdullah University of Science and Technology\\
	Thuwal 23955-6900, Saudi Arabia \\
        \And
    {Tariq Alkhalifah} \\
	Division of Physical Science and Engineering\\
	King Abdullah University of Science and Technology\\
	Thuwal 23955-6900, Saudi Arabia \\
    [3ex]
  $^{*}$Corresponding author: \textbf{Shijun Cheng}~(\texttt{sjcheng.academic@gmail.com})
}

\renewcommand{\shorttitle}{Generative wave propagator}

\hypersetup{
pdftitle={A template for the arxiv style},
pdfsubject={q-bio.NC, q-bio.QM},
pdfauthor={David S.~Hippocampus, Elias D.~Striatum},
pdfkeywords={First keyword, Second keyword, More},
}

\begin{document}
\maketitle

\begin{abstract}
Seismic wavefield simulation is fundamental to seismology, but conventional finite-difference (FD) methods remain limited by numerical dispersion and stability constraints, which often require dense spatial grids and small time steps and thereby severely limit the effectiveness of iterative inversion workflows. We introduce a conditional diffusion-based wavefield propagator that advances seismic wavefields recursively from one time step to the next. Instead of learning an unconditional data distribution of wavefield evolution, the model is conditioned by a short history of recent wavefield time steps (snapshots), the velocity model, and the wavefield time step index, allowing it to represent the conditional transition between adjacent physical states. By training the network to directly predict the clean next wavefield snapshot, this strong physical conditioning makes it possible to replace the iterative reverse diffusion process with a single network evaluation for each predicted snapshot. To improve stability over long recursive rollouts, we further introduce a causal time-weighted loss, in which adaptive weights, accumulated as exponential moving averages of per-snapshot training errors, emphasize training directions that are consistent with the forward propagation sequence and reduce the amplification of one-step prediction errors. Because the learned propagator is tied to the temporal spacing of the training snapshots rather than to the FD stability limit, it can advance the wavefield using a physical time step ten times larger than that required by the underlying solver. Experiments on the Overthrust, SEG/EAGE, and Marmousi models show that the proposed method accurately reproduces wavefield snapshots and shot gathers for in-distribution structures, preserves wavefront geometry and traveltimes under geological distribution shift, and achieves an end-to-end speedup of $2.17\times$ over a GPU-accelerated tenth-order staggered-grid FD implementation under matched hardware conditions.
\end{abstract}

\keywords{Wave propagation \and Generative diffusion model}
\section{\textbf{Introduction}}
Seismic wavefield simulation is a fundamental problem in the seismological community, serving as the cornerstone for numerous applications in exploration geophysics and earthquake seismology \citep{komatitsch1999introduction, virieux2009overview}. Accurate forward modeling of seismic wave propagation is essential for understanding subsurface structures, designing acquisition geometries, interpreting seismic data, and implementing imaging and inversion workflows \citep{tarantola1984inversion, fichtner2010full}. The ability to efficiently simulate realistic wavefields directly impacts the success of hydrocarbon exploration, geothermal resource assessment, and seismic hazard evaluation. 

Traditional numerical methods for seismic wavefield simulation have been extensively developed and widely adopted over the past decades. Finite-difference methods (FDMs) \citep{virieux1984sh, virieux1986p, levander1988fourth, moczo2007finite, liu2013globally, cheng2021wave}, finite-element methods (FEMs) \citep{marfurt1984accuracy, komatitsch1999introduction, komatitsch2002spectral}, and pseudo-spectral methods (PSMs) \citep{kosloff1982forward, fornberg1987pseudospectral, reshef1988three, zhu2014modeling, wang2022propagating} have demonstrated their effectiveness in modeling wave propagation through complex geological models. These methods have been successfully applied to wave equations, including those for acoustic, elastic, anisotropic, and poroelastic media \citep{carcione1996wave, alkhalifah2000acoustic, moczo2007finite}. However, these conventional approaches suffer from inherent limitations that constrain their practical application. For example, FDMs suffer from numerical dispersion artifacts, requiring careful selection of grid spacing, time step, and dominant frequency to achieve accurate wavefield simulation \citep{levander1988fourth, fei1995elimination}. Fine-grained simulations with dispersion-free wavefields demand dense spatial sampling and relatively low dominant frequencies, which significantly increase the number of grid points for a given model size and consequently impose substantial computational burdens \citep{moczo2007finite, virieux2009overview}. Furthermore, stability conditions restrict the maximum allowable time step size to prevent numerical instability, forcing the use of small time steps that dramatically increase the computational cost of forward modeling \citep{moczo2007finite}. This computational expense becomes a particular bottleneck for iterative inversion workflows, where hundreds of wavefield simulations are commonly required \citep{virieux2009overview, alkhalifah2016full}.

To address the limitations of traditional numerical solvers, recent efforts have explored machine learning approaches as alternative wavefield simulators. Physics-informed neural networks (PINNs) have been used to solve the Helmholtz equation for frequency-domain scattered wavefield modeling \citep{alkhalifah2021wavefield, song2021solving, rasht2022physics}, offering the advantage of mesh-free computation and automatic differentiation. However, PINNs serve as function approximators that must be retrained for each individual velocity model and frequency, limiting their practical efficiency \citep{cheng2025meta, cheng2026multifrequency}. To overcome this single-model limitation, operator learning frameworks have been proposed to generalize across different velocity models \citep{yang2021seismic}. \cite{huang2025learned} used Fourier neural operators (FNO) for frequency-domain scattered wavefield simulation, enabling rapid prediction for unseen velocity models. \cite{cheng2025seismic} further advanced this direction by proposing generative neural operators to enhance the generalization capability beyond Huang et al.'s work. In the time domain, \cite{yang2021seismic} and \cite{zhang2023learning} adapted FNOs to simulate seismic wavefields by taking velocity models as input and producing a series of wavefield snapshots at multiple time steps as output. However, this one-shot generation strategy, which simultaneously produces all temporal snapshots, faces significant challenges when generalizing to out-of-distribution velocity models, particularly for complex geological structures that differ substantially from the training data.

In this study, we propose a novel approach for generative wavefield simulation using conditional diffusion models. Instead of generating all time steps simultaneously, we leverage the progressive nature of wavefield evolution and design a conditional diffusion framework that learns the physics of wave propagation in a recursive manner. Specifically, we train a conditional diffusion model in which the target distribution is the wavefield at the next time step, conditioned on a short history of recent wavefield time steps (snapshots), the velocity model, and the wavefield time-step index. During inference, starting from any given initial wavefield (e.g., a source signature), we recursively apply the learned diffusion model to advance the wavefield through time, with each step requiring only a single forward pass of the network. To suppress the accumulation of one-step prediction errors that is inherent to recursive inference, we further introduce a causal time-weighted loss whose adaptive weights, accumulated across training iterations, align the training direction with the physical direction of wave propagation: the network is encouraged to learn snapshots near the source first, and later snapshots only after the earlier ones have been fitted to acceptable accuracy. This recursive generation strategy, combined with the causal training scheme, offers several key advantages:
\begin{itemize}
    \item It naturally captures the temporal causality of wave propagation;
    \item It relaxes the stability constraint inherent in traditional numerical methods, enabling large time-step simulations;
    \item It decomposes the challenging problem of long-horizon prediction into manageable single-step transitions, with error propagation explicitly controlled by training-time weighting;
    \item The diffusion single-step inference exploits the strong physical conditioning of the problem, eliminating the iterative reverse sampling required by standard diffusion models.
\end{itemize}
We test the effectiveness of our proposed propagator on the Overthrust, SEG/EAGE, and Marmousi models against finite-difference (FD) references on in-distribution geology, as well as out-of-distribution examples.
\section{\textbf{Method}}\label{sec:method}

\subsection{Problem statement}\label{sec:method:sub1}

Seismic wave propagation in acoustic isotropic media with constant density is governed by a second-order wave equation
\begin{equation}\label{eq:wave}
\frac{1}{v^2(\mathbf{x})}\frac{\partial^2 u(\mathbf{x},t)}{\partial t^2} = \nabla^2 u(\mathbf{x},t) + s(\mathbf{x},t),
\end{equation}
where $u(\mathbf{x},t)$ denotes the pressure wavefield at spatial location $\mathbf{x}$ and time $t$, $v(\mathbf{x})$ is the spatially varying velocity model, $s(\mathbf{x},t)$ is the source term, and $\nabla^2$ is the Laplacian operator. The dominant numerical approach for solving equation~(\ref{eq:wave}) is FDM, which discretizes both spatial derivatives and the temporal evolution on a regular grid. With an explicit second-order time-stepping scheme, the wavefield is advanced as
\begin{equation}\label{eq:fdm}
u^{n+1} = F\!\left(u^{n},\, u^{n-1},\, v,\, \Delta t_{\text{FD}}\right),
\end{equation}
where $u^{n}$ is the wavefield at the $n$-th wavefield time step, $\Delta t_{\text{FD}}$ is the FDM time interval, and $F$ is the discrete wave-equation operator determined by the chosen FD stencil, and can represent any wave equation.

Although equation~(\ref{eq:fdm}) is conceptually simple, its computational cost is fundamentally constrained by two discretization conditions. To suppress numerical dispersion, the spatial grid spacing must satisfy \citep{moczo2014finite}
\begin{equation}\label{eq:dispersion}
\Delta x \;\leq\; \frac{v_{\min}}{N_{\lambda}\, f_{\max}},
\end{equation}
where $v_{\min}$ is the minimum velocity in the medium, $f_{\max}$ is the maximum frequency of interest, and $N_{\lambda}$ is the required number of grid points per wavelength, typically between five and ten depending on the stencil order. To ensure stability of the explicit time integration, the time step must in turn satisfy the Courant--Friedrichs--Lewy (CFL) condition \citep{levander1988fourth}
\begin{equation}\label{eq:cfl}
\Delta t_{\text{FD}} \;\leq\; C \cdot \frac{[\Delta x, \Delta y, \Delta z]_\text{min}}{v_{\max}},
\end{equation}
where $v_{\max}$ is the maximum velocity, $[\Delta x, \Delta y, \Delta z]_\text{min}$ denotes the minimum spatial grid spacing, and $C$ is a scheme-dependent constant of order unity.

Two observations about equations~(\ref{eq:dispersion}) and~(\ref{eq:cfl}) motivate the present work. First, the two constraints are coupled and compounding: a finer spatial grid (driven by equation~(\ref{eq:dispersion})) forces a proportionally smaller time step (through equation~(\ref{eq:cfl})), so the total number of stencil evaluations required to propagate a wavefield to a fixed physical time $T$ scales unfavorably with $f_{\max}$ and with $v_{\max}/v_{\min}$. Second, neither constraint is a property of the underlying physics; both arise solely from the act of discretizing equation~(\ref{eq:wave}). In iterative inversion workflows such as full waveform inversion (FWI) and reverse-time migration (RTM), where the forward problem must be solved thousands of times, this overhead dominates the overall computational budget.

These observations suggest an alternative formulation in which the wavefield evolution is treated not as the solution of a discretized PDE, but as the action of a propagation operator that can be learned directly from data. Specifically, we seek an operator
\begin{equation}\label{eq:operator}
\mathcal{P}_{\Delta t}: \; \bigl(u^{n},\, v\bigr) \;\longmapsto\; u^{n+1},
\end{equation}
that advances the wavefield by a physical time increment $\Delta t$. Crucially, because $\mathcal{P}_{\Delta t}$ is learned from samples rather than derived from an FD stencil, its effective $\Delta t$ is no longer bound by equation~(\ref{eq:cfl}). Therefore, it is bounded only by the temporal resolution of the training data and by the expressiveness of the model. The remainder of this section develops $\mathcal{P}_{\Delta t}$ as a conditional generative diffusion model.

\subsection{Background: generative diffusion models}\label{sec:method:sub2}

Generative diffusion models (GDMs) define a generative process by reversing a gradual noising of data \citep{ho2020denoising}. They consist of two coupled stochastic processes: a fixed forward process that progressively corrupts a clean sample $\mathbf{x}_0$ into pure noise, and a learned reverse process that recovers $\mathbf{x}_0$ from noise.

The forward process is a Markov chain of $T$ steps that injects Gaussian noise according to a prescribed variance schedule $\{\beta_t\}_{t=1}^{T}$ resulting in the following conditional distribution:
\begin{equation}\label{eq:forward}
q(\mathbf{x}_t \mid \mathbf{x}_{t-1}) \;=\; \mathcal{N}\!\left(\mathbf{x}_t;\, \sqrt{1-\beta_t}\,\mathbf{x}_{t-1},\; \beta_t\,\mathbf{I}\right).
\end{equation}
A standard property of this chain is that the marginal distribution of $\mathbf{x}_t$ given $\mathbf{x}_0$ is available in closed form, which allows direct sampling at any noise level without iterating through intermediate steps:
\begin{equation}\label{eq:marginal}
\mathbf{x}_t \;=\; \sqrt{\bar{\alpha}_t}\,\mathbf{x}_0 \;+\; \sqrt{1-\bar{\alpha}_t}\,\boldsymbol{\epsilon}, \qquad \boldsymbol{\epsilon} \sim \mathcal{N}(\mathbf{0},\, \mathbf{I}),
\end{equation}
where $\alpha_t = 1 - \beta_t$ and $\bar{\alpha}_t = \prod_{i=1}^{t} \alpha_i$. As $t$ increases, $\bar{\alpha}_t$ decays from near one toward zero, and the distribution of $\mathbf{x}_t$ slowly evolves from the data distribution to an isotropic Gaussian.

The reverse process is parameterized as a Markov chain in the opposite direction, with Gaussian transitions whose mean is predicted by a neural network with parameters $\theta$, and represented by the following distribution:
\begin{equation}\label{eq:reverse}
p_\theta(\mathbf{x}_{t-1} \mid \mathbf{x}_t) \;=\; \mathcal{N}\!\left(\mathbf{x}_{t-1};\, \boldsymbol{\mu}_\theta(\mathbf{x}_t, t),\; \tilde{\beta}_t\,\mathbf{I}\right).
\end{equation}
Training proceeds by maximizing a variational lower bound on the data log-likelihood, which reduces, after standard simplifications \citep{ho2020denoising}, to a denoising regression objective on samples drawn at random noise levels $t \sim \text{Uniform}(1, T)$.

A subtle but practically important degree of freedom in this framework lies in how the regression target is parameterized. Three choices are common in the literature. The first, $\epsilon$-prediction, trains the network to predict the noise $\boldsymbol{\epsilon}$ that was added in equation~(\ref{eq:marginal}) \citep{ho2020denoising}. This is the original formulation and is well suited for natural images, where the clean signal occupies a complex, multi-modal manifold. The second, $\mathbf{x}_0$-prediction, trains the network to directly predict the clean sample $\mathbf{x}_0$ from its noisy version $\mathbf{x}_t$ \citep{bansal2024cold}. The third predicts a linear combination of the two and is primarily motivated by improved behavior at the noise-schedule endpoints \citep{zhang2023improving}. The three parameterizations are mathematically equivalent in the sense that any one can be converted into the other given $\mathbf{x}_t$ and the schedule $\{\bar{\alpha}_t\}$, but they induce different loss landscapes and different empirical behavior at inference time, especially when the number of sampling steps is reduced.

In the unconditional formulation above, the reverse process generates samples from the marginal data distribution $p(\mathbf{x}_0)$. For our application, however, the quantity of interest is not the marginal distribution of wavefields but the conditional distribution of the next wavefield snapshot given the current state of the simulation. Conditional diffusion models extend equations~(\ref{eq:forward})-(\ref{eq:reverse}) by allowing the network to take auxiliary inputs $\mathbf{c}$, yielding a learned reverse process $p_\theta(\mathbf{x}_0 \mid \mathbf{c})$. The forward process is unchanged, and only the denoiser is conditioned. In the next section, we instantiate this framework with $\mathbf{c}$ comprising a short history of recent wavefield time steps (snapshots), the velocity model, and the wavefield time-step index, and we adopt the $\mathbf{x}_0$-prediction parameterization, deferring the justification of this choice to Section~\ref{sec:method:sub3:sub2}.

\subsection{Conditional diffusion model for wavefield propagation}\label{sec:method:sub3}

\subsubsection{Conditional formulation}\label{sec:method:sub3:sub1}

We cast wavefield propagation as a conditional generative modeling task: learning the conditional distribution $p_\theta(u^{n+1} \mid \mathbf{u}^{n-4:n}, v, n)$ of the next wavefield snapshot given a short history of recent wavefield snapshots (in our case 5), the velocity model $v$, and the wavefield time-step index $n$. Here, $\mathbf{u}^{n-4:n} = \{u^{n-4}, u^{n-3}, u^{n-2}, u^{n-1}, u^{n}\}$, which is a set of the five most recent snapshots, are passed to the network as a multi-channel input. Conditioning on a short history rather than on the single most recent snapshot provides the network with implicit access to short-range temporal gradients of the wavefield, which encode information about the local direction and energy of wave propagation that is not contained in any single instant in isolation. The history length is a design choice. We use five frames throughout this work, a practical setting that we find sufficient to encode the local temporal structure of the wavefield at the time step $\Delta t$ at which our model operates, without significantly increasing the input dimensionality of the network. For snapshots near the source where fewer than five preceding frames are physically defined, the missing entries of $\mathbf{u}^{n-4:n}$ are filled with zeros, reflecting the physical fact that the wavefield is identically zero prior to source excitation.

Compared to the unconditional formulation of Section~\ref{sec:method:sub2}, the three conditioning components encode complementary information. The history $\mathbf{u}^{n-4:n}$ provides the causal state from which the next snapshot should evolve, carrying both the present configuration and the recent dynamics of all waves currently active in the domain. The velocity model $v$ specifies the medium and therefore determines, together with $\mathbf{u}^{n-4:n}$, the local geometry of propagation, reflection, and refraction over the time increment $\Delta t$. The wavefield time-step index $n$ is supplied through a sinusoidal positional embedding and allows the propagation operator to adapt to time-dependent characteristics of the wavefield sequence.

Within the diffusion framework, the forward process of Section~\ref{sec:method:sub2} is applied to the target snapshot $u^{n+1}$, producing a noised version $\mathbf{x}_t = \sqrt{\bar{\alpha}_t}\,u^{n+1} + \sqrt{1-\bar{\alpha}_t}\,\boldsymbol{\epsilon}$ at diffusion step $t$. The reverse process is then conditioned on $(\mathbf{u}^{n-4:n}, v, n)$ in addition to $t$. Denoting the conditional denoiser by $f_\theta$, the base training objective is the conditional analogue of the standard denoising loss,
\begin{equation}\label{eq:base_loss}
\mathcal{L}_{\text{base}} \;=\; \mathbb{E}_{n,\,t,\,u^{n+1},\,\boldsymbol{\epsilon}}\!\left[\,\bigl\|u^{n+1} - f_\theta(\mathbf{x}_t,\, t,\, \mathbf{u}^{n-4:n},\, v,\, n)\bigr\|^{2}\right],
\end{equation}
where $\mathbf{x}_t \;=\; \sqrt{\bar{\alpha}_t}\,u^{n+1} \;+\; \sqrt{1-\bar{\alpha}_t}\,\boldsymbol{\epsilon}$ (i.e., using equation (\ref{eq:marginal})). Two refinements of equation~(\ref{eq:base_loss}) are developed in the following subsections: the parameterization of the regression target (Section~\ref{sec:method:sub3:sub2}) and a causal time-weighting scheme that addresses error propagation under recursive inference (Section~\ref{sec:method:sub3:sub3}).

\subsubsection{Direct $\mathbf{x}_0$-prediction parameterization}\label{sec:method:sub3:sub2}

As reviewed in Section~\ref{sec:method:sub2}, the diffusion denoiser can be parameterized to predict either the injected noise $\boldsymbol{\epsilon}$, the clean signal $\mathbf{x}_0$, or a linear combination of the two. We adopt, as reflected by equation (\ref{eq:base_loss}), the $\mathbf{x}_0$-prediction parameterization, so that $f_\theta$ directly outputs the clean target wavefield $u^{n+1}$ given the noisy input $\mathbf{x}_t$. Two considerations motivate this choice.

First, seismic wavefields are highly structured signals: they exhibit strong local coherence along wavefronts, sparse spatial support away from the active wavefront, and smooth amplitude variation governed by the medium. The $\mathbf{x}_0$-prediction objective regresses directly onto this structure, whereas $\boldsymbol{\epsilon}$-prediction regresses onto isotropic Gaussian noise that has been linearly mixed with the signal. For wavefield data, the former yields a loss landscape more aligned with the geometry of the target manifold and produces more uniform behavior across noise levels.

Second, $\mathbf{x}_0$-prediction is naturally compatible with the one-step sampling strategy described in Section~\ref{sec:method:sub4}. Because $f_\theta$ is trained to recover $u^{n+1}$ from $\mathbf{x}_t$ at all noise levels $t \in [1, T]$, it learns a family of denoisers mapping arbitrary noise scales onto the clean wavefield. At inference time, this allows the most extreme noise level to be exploited without iterating through the reverse chain, an option that is far less natural under the $\boldsymbol{\epsilon}$ parameterization, where the network would output noise rather than signal.

\subsubsection{Causal time-weighted loss}\label{sec:method:sub3:sub3}

The base objective of equation~(\ref{eq:base_loss}) treats all snapshot indices $n$ on equal footing. This is appropriate when each $(\mathbf{u}^{n-4:n}, u^{n+1})$ pair is regarded as an independent training sample, but it ignores a fundamental asymmetry of recursive inference: a prediction error introduced at snapshot $n$ is propagated forward into every subsequent snapshot $n+1, n+2, \ldots, N-1$, because each step consumes the previous prediction as part of its conditioning input. Errors at early snapshots therefore have a disproportionate impact on long-horizon simulation accuracy, while errors at the final few snapshots barely influence the rest of the sequence. A uniformly weighted loss can encourage the network to fit late, comparatively easy snapshots at the expense of the early, more consequential ones.

A natural remedy is to weight the per-sample loss by wavefield time-step index, suppressing the contribution of later snapshots until earlier ones have been well learned. A fixed schedule, however, is inflexible: it presupposes which snapshots are difficult before training has begun, and cannot adapt to the evolution of the network's per-snapshot competence over the course of training. We therefore adopt an adaptive scheme inspired by the causal training strategy proposed for PINNs by \cite{wang2022respecting}, in which the weight of a given temporal index is suppressed in proportion to the cumulative running loss at earlier indices, so that an index is admitted into the active training set only after the earlier snapshots have been learned to acceptable accuracy \citep{huang2024microseismic}.

Specifically, for each wavefield time-step index $n \in \{0, 1, \ldots, N-1\}$ we maintain a running estimate $L_{\text{ema}}[n]$ of the per-snapshot prediction loss, updated by an exponential moving average across training iterations,
\begin{equation}\label{eq:ema_update}
L_{\text{ema}}^{(i+1)}[n] \;=\; \gamma\, L_{\text{ema}}^{(i)}[n] \;+\; (1-\gamma)\, \bigl\|u^{n+1} - f_\theta(\mathbf{x}_t,\, t,\, \mathbf{u}^{n-4:n},\, v,\, n)\bigr\|^{2},
\end{equation}
where $i$ indexes the training iteration and $\gamma \in [0, 1]$ is the exponential moving average (EMA) decay rate. The update in equation~(\ref{eq:ema_update}) is applied only to those snapshot indices that appear in the current minibatch, and the remaining entries of $L_{\text{ema}}$ are left unchanged. This selective update decouples the running estimate from the stochasticity of minibatch sampling and lets $L_{\text{ema}}$ accumulate information across the entire training history. The vector is initialized to all ones, encoding the prior assumption that all snapshots are equally untrained at the start of training.

Given $L_{\text{ema}}$, the causal weight associated with snapshot $n$ is defined as
\begin{equation}\label{eq:omega}
\omega(n) \;=\;
\begin{cases}
1, & \text{if } \exp\!\left(-\varepsilon \sum_{k=0}^{n-1} L_{\text{ema}}[k]\right) \,\geq\, \delta, \\[4pt]
\exp\!\left(-\varepsilon \sum_{k=0}^{n-1} L_{\text{ema}}[k]\right), & \text{otherwise},
\end{cases}
\end{equation}
where $\varepsilon > 0$ controls the rate of causal decay and $\delta \in (0, 1)$ is a saturation threshold. The cumulative sum $\sum_{k=0}^{n-1} L_{\text{ema}}[k]$ quantifies the total residual error at all snapshots strictly preceding $n$. When this quantity is large, $\omega(n)$ is small and snapshot $n$ contributes negligibly to the gradient. As training progresses and the early entries of $L_{\text{ema}}$ shrink, the cumulative sum decreases, $\omega(n)$ grows, and progressively later snapshots are admitted into the active training set. The threshold $\delta$ collapses values of $\omega(n)$ that are already very close to one onto exactly one, both to prevent numerical drift near saturation and to provide a clean convergence indicator: training is regarded as having propagated through the entire wavefield sequence once $\min_{n}\,\omega(n) \,\geq\, \delta$.

Combining the causal weight with the base loss of equation~(\ref{eq:base_loss}) yields the full training objective
\begin{equation}\label{eq:causal_loss}
\mathcal{L} \;=\; \mathbb{E}_{n,\,t,\,u^{n+1},\,\boldsymbol{\epsilon}}\!\left[\,\omega(n)\, \bigl\|u^{n+1} - f_\theta(\mathbf{x}_t,\, t,\, \mathbf{u}^{n-4:n},\, v,\, n)\bigr\|^{2}\right],
\end{equation}
in which $\omega(n)$ is treated as a constant with respect to the model parameters at each gradient step, since it is computed from the stop-gradient EMA buffer $L_{\text{ema}}$. The resulting procedure respects the causal structure of the wavefield-evolution problem: the network is encouraged to first learn an accurate one-step propagator for snapshots near the source, and only then to refine its predictions at progressively later times. We report the empirical effect of this scheme on training dynamics and on long-horizon accuracy in Section~\ref{sec:examples}.

\subsubsection{Source injection and boundary handling}\label{sec:method:sub3:sub4}

The source term $s(\mathbf{x}, t)$ of the governing wave equation (equation~(\ref{eq:wave})) and the treatment of the simulation-domain boundary are both handled implicitly within our formulation. The source is encoded through the initial wavefield $u^{0}$, which is set to the source wavelet localized at the shot position, with all conditioning frames preceding $u^{0}$ taken to be zero as noted in Section~\ref{sec:method:sub3:sub1}. Because the network is trained on data produced under the same convention, no additional source term need be injected at subsequent snapshots, and the learned operator $\mathcal{P}_{\Delta t}$ advances the source-imprinted wavefield in the same manner as any other state. The treatment of the simulation-domain boundary is likewise inherited from the training data: whatever absorbing-boundary scheme is applied by the FD solver during data generation (e.g., a perfectly matched layer) is learned by the trained operator near the domain edges through statistical matching to the training distribution. Explicit boundary conditions therefore need not be imposed at inference time, and the propagation operator defined in equation~(\ref{eq:operator}) is complete in the sense that it requires no auxiliary information beyond $(\mathbf{u}^{n-4:n}, v)$.

\subsection{Recursive inference with one-step sampling}\label{sec:method:sub4}

Standard inference in a diffusion model follows the reverse Markov chain of equation~(\ref{eq:reverse}), iterating from a pure-noise initialization at diffusion step $t=T$ down to $t=0$ over a sequence of $T$ denoising updates. Applied directly to our conditional model, this would require $T$ forward passes of $f_\theta$ per wavefield snapshot, and, since a full simulation comprises $N$ snapshots advanced one after another, $T \times N$ forward passes in total. For values of $T$ commonly used in practice, typically several hundred to a few thousand, this iterative cost is incompatible with the goal of accelerating forward modeling relative to FDM.

We exploit the $\mathbf{x}_0$-prediction parameterization of Section~\ref{sec:method:sub3:sub2} to collapse the reverse chain into a single prediction (reverse) pass. Recall that $f_\theta$ is trained at diffusion steps $t$ drawn uniformly from $[1, T]$ to recover the clean target $u^{n+1}$ from a noisy input $\mathbf{x}_t = \sqrt{\bar{\alpha}_t}\,u^{n+1} + \sqrt{1-\bar{\alpha}_t}\,\boldsymbol{\epsilon}$, conditioned on $(\mathbf{u}^{n-4:n}, v, n)$. Because the conditioning strongly constrains the prediction, the conditional distribution $p_\theta(u^{n+1} \mid \mathbf{u}^{n-4:n}, v, n)$ has low residual entropy, and a sufficiently expressive network learns to recover the target accurately across the full range of training noise levels. The noisy input $\mathbf{x}_t$ therefore acts during training primarily as a stochastic regularizer rather than as a carrier of information about $u^{n+1}$, which the network must in any case reconstruct from the conditioning channels.

Building on this observation, at inference time we replace the full reverse chain by the single forward pass
\begin{equation}\label{eq:one_step}
\hat{u}^{n+1} \;=\; f_\theta(\mathbf{z},\, t=T,\, \mathbf{u}^{n-4:n},\, v,\, n+1), \qquad \mathbf{z} \sim \mathcal{N}(\mathbf{0},\, \mathbf{I}),
\end{equation}
in which the noisy input is supplied as a fresh draw of isotropic Gaussian noise $\mathbf{z}$ and the diffusion-step embedding is fixed to $t=T$. The choice $t=T$ is consistent with the $\mathbf{x}_0$-prediction parameterization, under which the network output is interpreted directly as a clean target rather than as a partially-denoised intermediate. The noise input $\mathbf{z}$ contributes residual stochasticity inherited from the training distribution, while the actual prediction is dominated by the conditioning channels $(\mathbf{u}^{n-4:n}, v, n+1)$. This procedure trades the iterative refinement of standard diffusion sampling for a single non-iterative single pass, in a spirit related to consistency-model and distillation-based fast samplers \citep{song2023consistency,salimans2022progressive}, though here the single-step capability arises from the strength of the physical conditioning rather than from a specialized training objective. The per-snapshot cost is thereby reduced from $T$ evaluations of $f_\theta$ to one, an acceleration of $T \times$ relative to standard denoising diffusion probabilistic model (DDPM) sampling at the same network capacity.

A full simulation is then obtained by applying equation~(\ref{eq:one_step}) recursively. Given an initial wavefield $u^{0}$ encoding the source as described in Section~\ref{sec:method:sub3:sub4} and a velocity model $v$, the procedure iterates over snapshot indices $n = 0, 1, \ldots, N-1$, maintaining a sliding buffer of the five most recent snapshots $\mathbf{u}^{n-4:n}$ (with zero padding at the beginning of the sequence), sampling a fresh noise vector $\mathbf{z} \sim \mathcal{N}(\mathbf{0}, \mathbf{I})$ at each step, and computing $u^{n+1} = f_\theta(\mathbf{z},\, t=T,\, \mathbf{u}^{n-4:n},\, v,\, n+1)$. Each new prediction is then appended to the buffer for use as conditioning at the next step. This recursive form is to be contrasted with a one-shot alternative in which a single model output produces the entire snapshot sequence in parallel. Three considerations favor the recursive choice. First, it preserves the causal structure of physical wave propagation: the prediction at snapshot $n+1$ depends explicitly and only on the immediately preceding states $\mathbf{u}^{n-4:n}$ and the medium $v$, mirroring the Markovian character of the underlying time-domain wave equation. Second, it controls error accumulation more gracefully than long-horizon direct prediction: at each step, the network solves a comparatively local one-step problem rather than a global trajectory-prediction problem, and the causal time-weighted loss of Section~\ref{sec:method:sub3:sub3} is designed specifically to suppress the propagation of these one-step errors along the recursion. Third, the simulation length $N$ is decoupled from the architecture of $f_\theta$, so that the same trained model can be deployed at arbitrary horizons without retraining, including horizons longer than any individual training trajectory.

\subsection{Network architecture}\label{sec:method:sub5}
The conditional denoiser $f_\theta$ is implemented using a U-Net adapted from the improved DDPM codebase \citep{nichol2021improved}. The backbone comprises four resolution stages with feature dimensions $64$, $128$, $256$, and $512$, arranged as an encoder--bottleneck--decoder with skip connections between matching stages. Each stage contains residual blocks built from group normalization, sigmoid linear unit (SiLU) activations, and $3 \times 3$ convolutions, with self-attention layers inserted at the two coarsest resolutions to capture long-range spatial coherence in the wavefield. The final output projection is zero-initialized, so that the network begins training as the identity mapping on its main residual path, which stabilizes the early optimization dynamics.

The four conditioning signals identified in Section~\ref{sec:method:sub3}, including the noisy diffusion input $\mathbf{x}_t$, the diffusion step $t$, the spatial conditioning $(\mathbf{u}^{n-4:n}, v)$, and the wavefield time-step index $n$, are injected into the U-Net through complementary mechanisms. The two scalar indices $t$ and $n$ are first mapped to sinusoidal positional encodings and then passed through a two-layer multilayer perceptron (MLP), yielding embedding vectors of dimension $256$. The spatial conditioning $(\mathbf{u}^{n-4:n}, v)$ is stacked channel-wise, forming a six-channel input that comprises the five most recent wavefield snapshots and the velocity model, and is then processed independently by a shallow convolutional stem. The stem produces a feature map at the base channel width, which is concatenated with $\mathbf{x}_t$ at the U-Net entry. This provides a direct, full-resolution physical context alongside the diffusion input.

Inside every residual block, the two embeddings act on separate streams. The diffusion-step embedding of $t$ modulates the main residual branch through a FiLM-style feature-wise affine transformation, where the embedding generates scale and shift parameters that control the noise-level dependence of the denoiser. In parallel, the spatial conditioning feature map is adaptively downsampled to match the resolution of the current stage, modulated by the snapshot-index embedding of $n$ through an analogous FiLM operation, and fused with the residual output by channel-wise concatenation followed by a $3 \times 3$ convolution. This factorization reflects the distinct roles of the two indices: $t$ acts on the diffusion variable as a noise-level signal, while $n$ acts on the physical conditioning as a temporal-context signal. Repeating this injection at every stage of the U-Net allows the network to access the full set of conditioning signals at every spatial scale, rather than relying on a single point of fusion at the input.

\section{\textbf{Numerical examples}}\label{sec:examples}

We now evaluate the proposed conditional diffusion model through a series of numerical experiments designed to assess both its accuracy as a forward-modeling operator and the contribution of its individual design choices. We first describe the training data and training configuration. We then present in-distribution results on two benchmark models, the Overthrust model and the SEAM Arid model, which share geological characteristics with the training set but are not seen during training. Next, we examine out-of-distribution generalization on the Marmousi model, whose structural style differs markedly from the training data. Finally, we analyze the effect of the causal time-weighted loss through a dedicated ablation study, comparing training dynamics and long-horizon accuracy with and without the proposed weighting scheme. Throughout, predicted wavefields are compared against FD references, and accuracy is reported both qualitatively through wavefield snapshots and shot-gather comparisons and quantitatively through error metrics.

\subsection{Training data and configuration}\label{sec:examples:sub1}

The training velocity models are drawn from the open-source dataset of \cite{cheng_2026_19790506}, which provides $256 \times 256$ velocity patches extracted from a collection of industry benchmark models, including BP1994, BP2004, Hess, Otway, Sigsbee, SEG/EAGE, Overthrust, and SEAM Arid. The dataset comprises 7871 velocity patches in total, spanning a wide range of geological structures. To reduce computational cost of generating the training set (wavefields), all patches are resized to $128 \times 128$ prior to simulation.

For each velocity model, we simulate five shot gathers with source positions randomly distributed along the surface. The forward simulations are performed with a tenth-order staggered-grid FDM solving the first-order velocity-stress acoustic wave equation. A perfectly matched layer (PML) is applied around the model boundaries to attenuate boundary reflections. The spatial grid spacing is set to $10$~m and the temporal sampling interval to $0.001$~s for a total of 1001 time samples. The source is a Ricker wavelet with a dominant frequency of $15$~Hz. For each shot gather, we keep a wavefield snapshot every $0.01$~s, yielding $101$ snapshots per simulation. Each pair of consecutive snapshots therefore corresponds to a physical time increment of $0.01$~s, and the trained operator learns to advance the wavefield at this step size, well beyond the time step dictated by the stability condition of the underlying FD solver.

The network is trained for $550000$ iterations with a batch size of $36$. We use an AdamW optimizer with a fixed learning rate of $1 \times 10^{-4}$. The diffusion process is configured with $T = 1000$ steps. To stabilize training, an exponential moving average of the model parameters is maintained with a decay rate of $0.999$. The entire training procedure takes approximately $95$~hours on a single NVIDIA A100 GPU.

Once trained, the model is used for forward simulation as described in Section~\ref{sec:method:sub4}: the initial wavefield $u^{0}$ is set to the Ricker wavelet at $t = 0$~s  at the location of the source, the preceding entries of the conditioning history $\mathbf{u}^{-4:0}$ are set to zero as described in Section~\ref{sec:method:sub3:sub1}, and the conditional diffusion model recursively generates the subsequent wavefield snapshots. All experiments reported below follow this inference procedure.

\begin{figure}[!htbp]
\centering
\includegraphics[width=0.5\textwidth]{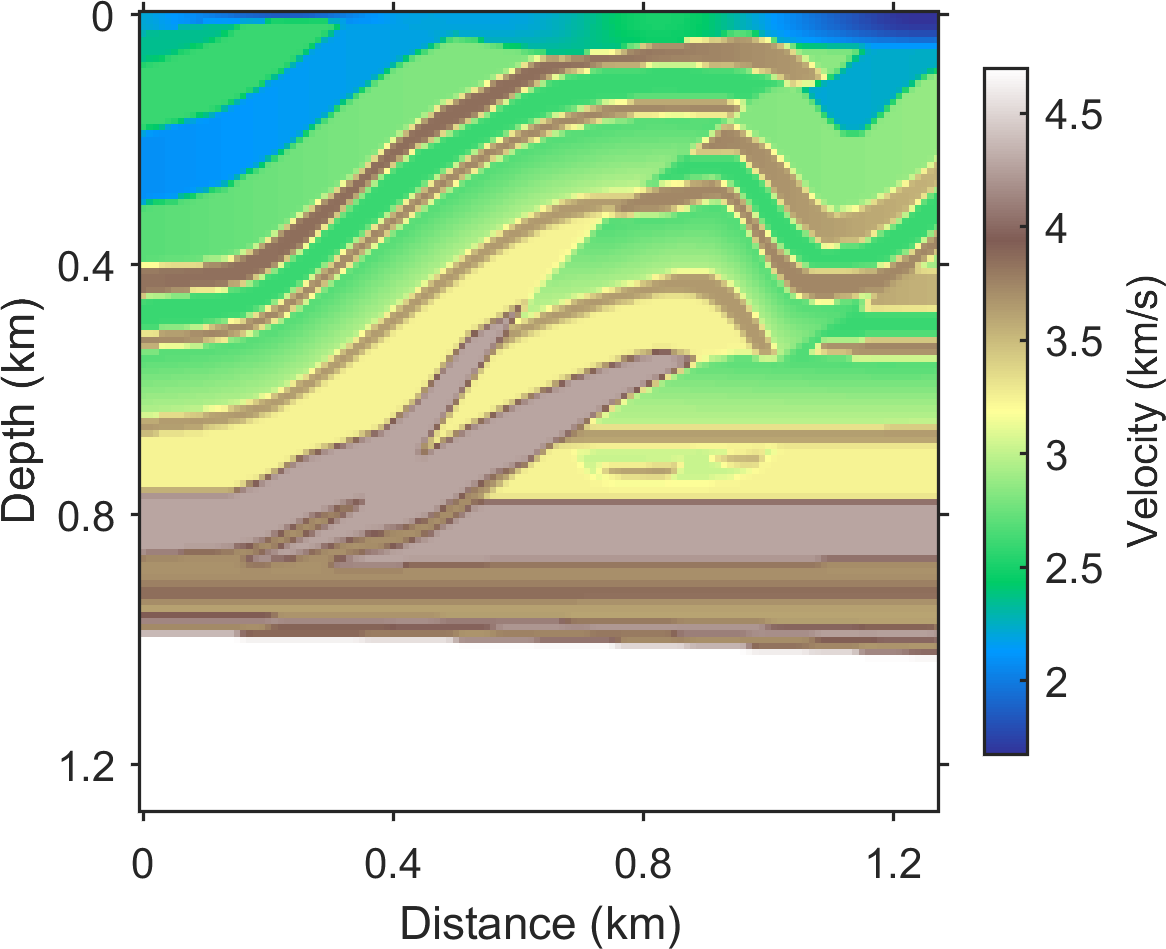}
\caption{The Overthrust velocity model used for in-distribution testing, resized to $128 \times 128$ with a spatial sampling of $10$~m.}
\label{fig1}
\end{figure}

\begin{figure}[!htbp]
\centering
\includegraphics[width=1\textwidth]{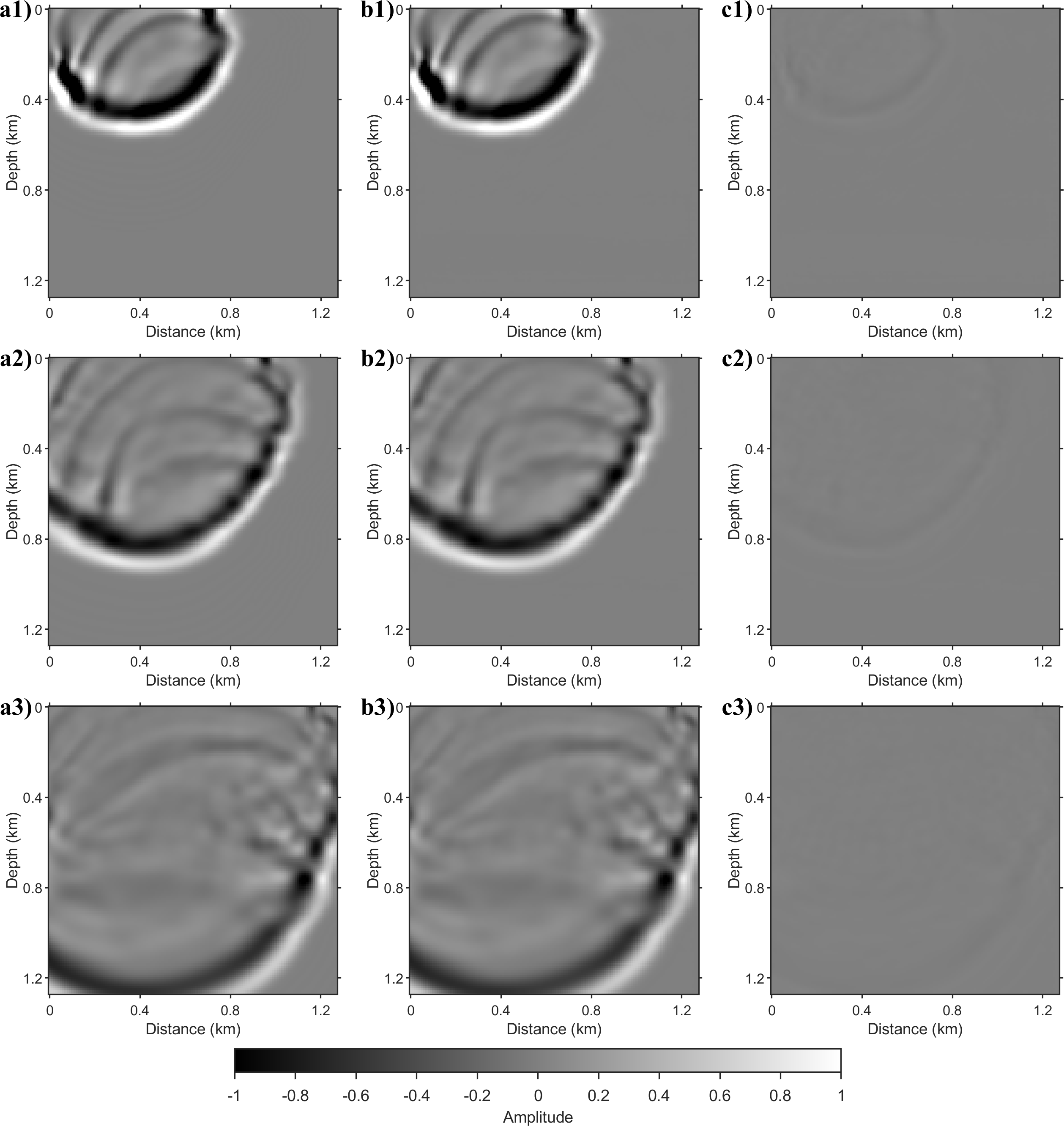}
\caption{Wavefield snapshot comparison for the Overthrust model with a source at $x = 0.32$~km on the surface. Rows from top to bottom correspond to the wavefields at $0.2$, $0.3$, and $0.4$~s. (a1)-(a3) FD reference snapshots. (b1)-(b3) Snapshots generated recursively by the proposed method starting from $t = 0$ s. (c1)-(c3) Difference between the reference and the prediction.}
\label{fig2}
\end{figure}

\begin{figure}[!htbp]
\centering
\includegraphics[width=1\textwidth]{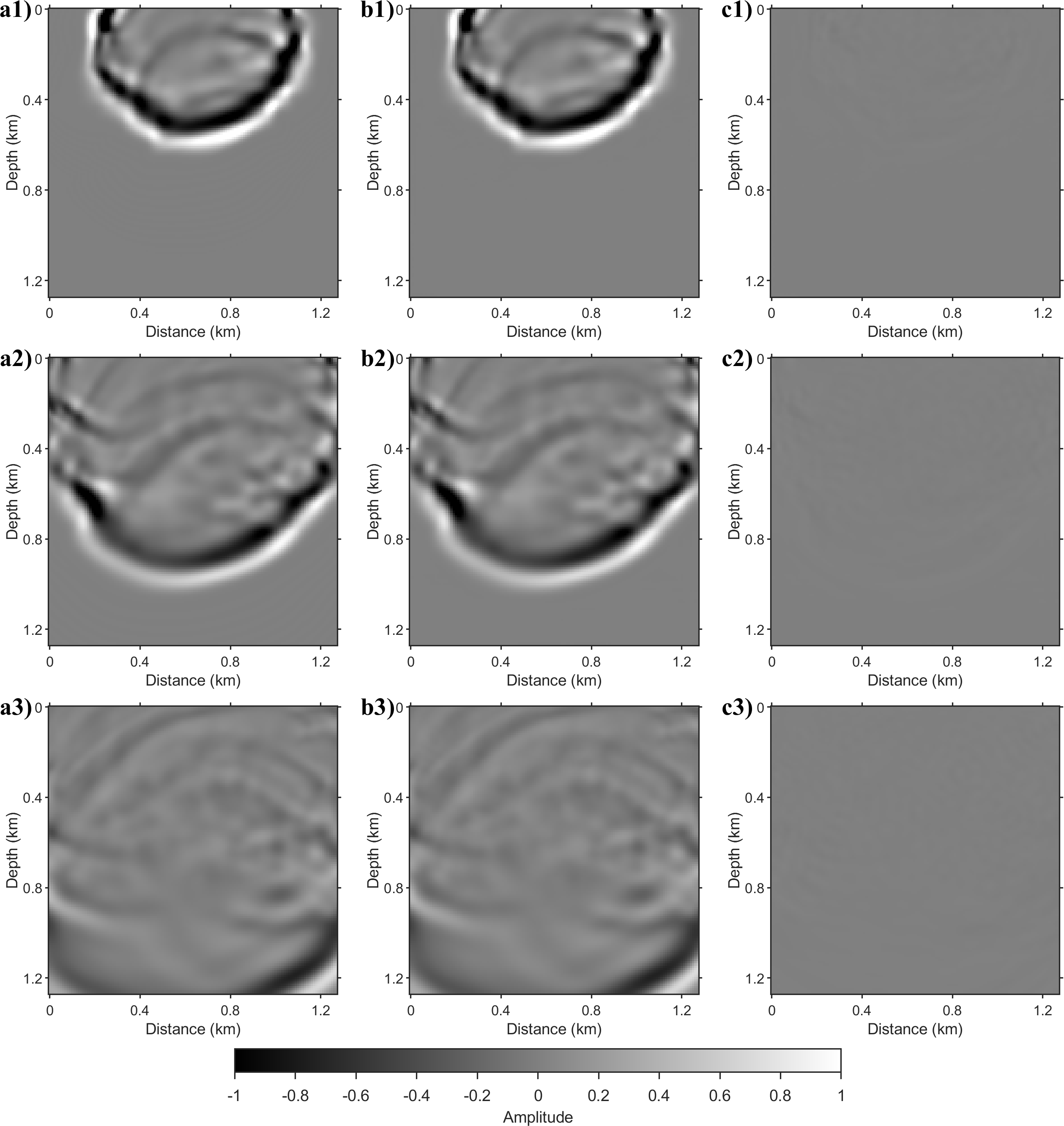}
\caption{Wavefield snapshot comparison for the Overthrust model with a source at $x = 0.64$~km on the surface, following the same arrangement as Figure~\ref{fig2}. }
\label{fig3}
\end{figure}

\begin{figure}[!htbp]
\centering
\includegraphics[width=0.95\textwidth]{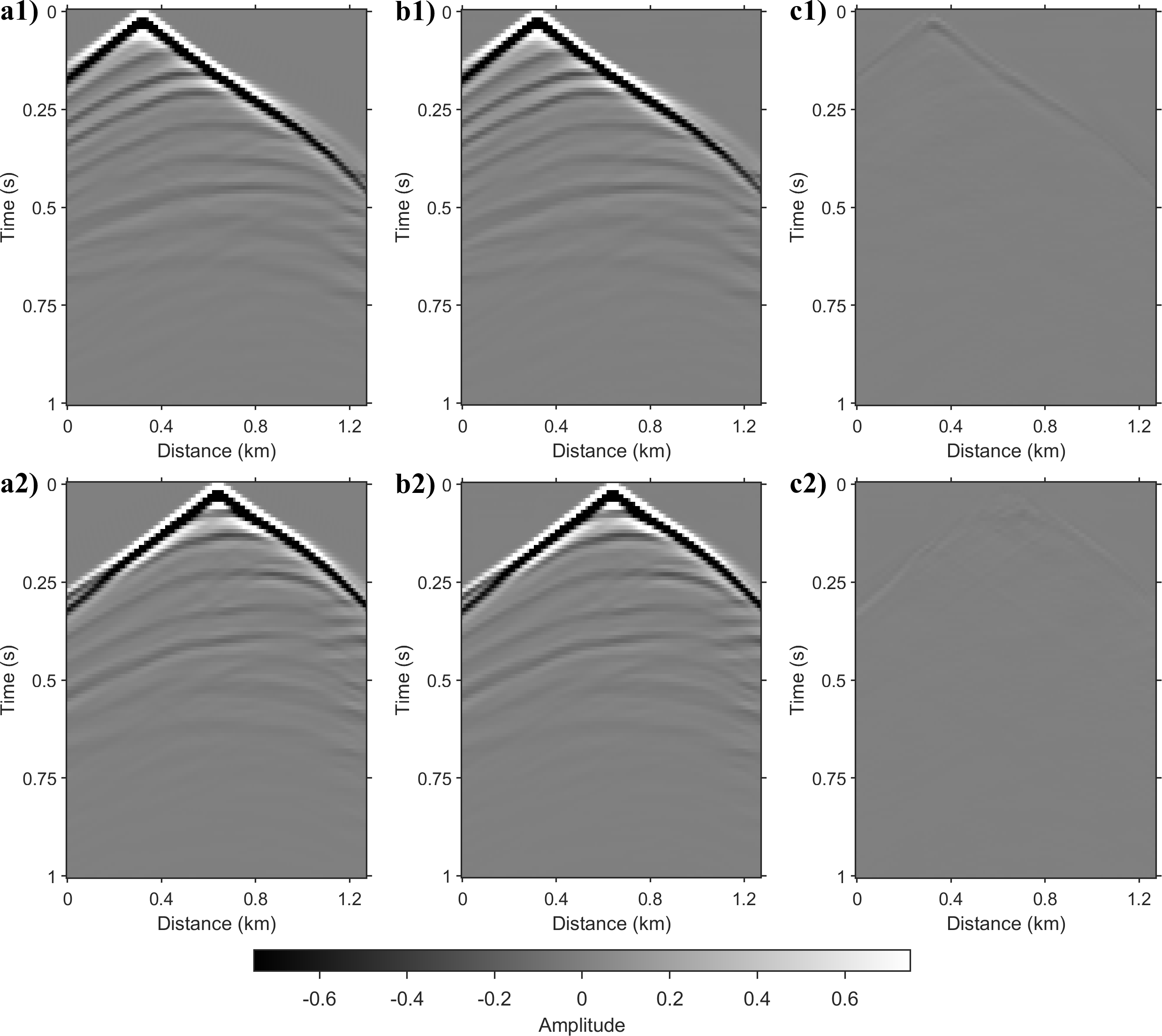}
\caption{Shot-gather comparison over $0$-$1$~s for the Overthrust model. Rows correspond to the sources at $x = 0.32$~km (top) and $x = 0.64$~km (bottom). (a1)-(a2) FD reference gathers. (b1)-(b2) Gathers generated by the recursive inference of the proposed method. (c1)-(c2) Difference between the reference and the prediction.}
\label{fig4}
\end{figure}

\begin{figure}[!htbp]
\centering
\includegraphics[width=0.95\textwidth]{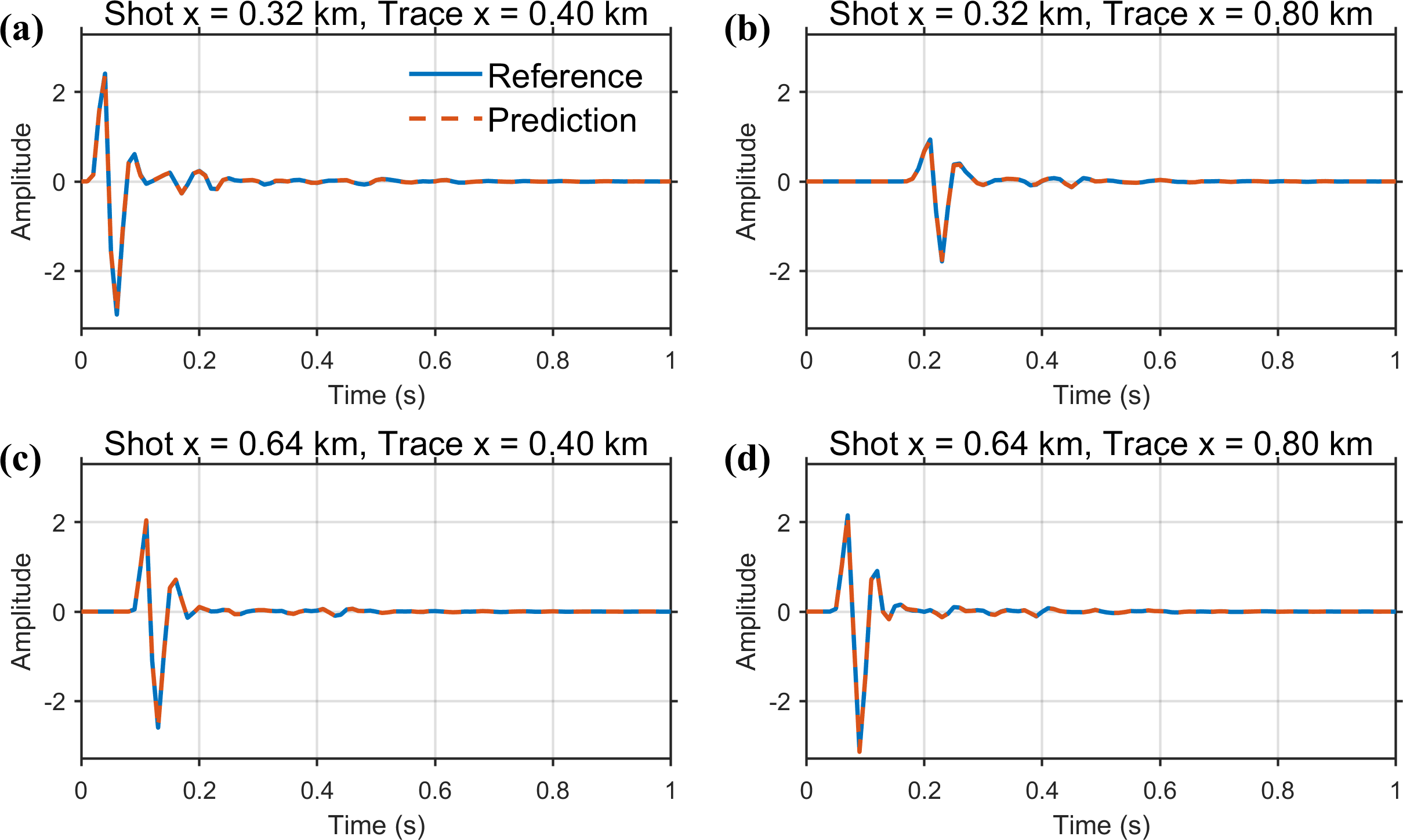}
\caption{Single-trace comparison for the Overthrust model. Traces are extracted at $x = 0.4$~km and $x = 0.8$~km from the shot gathers of the two sources: the top row corresponds to the source at $x = 0.32$~km and the bottom row to the source at $x = 0.64$~km. Solid blue lines denote the FD reference and dashed orange lines denote the prediction of the proposed method.}
\label{fig5}
\end{figure}

\subsection{In-distribution test: the Overthrust model}\label{sec:examples:sub2}

We begin the evaluation with the Overthrust model, a test that probes the behavior of the learned operator on geology well represented in the training distribution. Figure~\ref{fig1} shows the Overthrust velocity model used for testing, resized to $128 \times 128$ with a spatial sampling of $10$~m to match the training configuration. This experiment therefore isolates the question of whether the conditional diffusion model reproduces wavefield propagation faithfully on familiar geology, before the more demanding out-of-distribution test.

We first examine wavefield snapshots for a source placed at $x = 0.32$~km on the surface. Figure~\ref{fig2} compares the wavefield snapshots at $0.2$, $0.3$, and $0.4$~s: the left column shows the FD reference, the middle column shows the prediction generated recursively by our method starting from $t = 0$, and the right column shows their difference. Across all three time instants, the predicted wavefields reproduce the reference in close detail, like the position, curvature, and amplitude of the principal wavefronts, are all recovered. The difference panels remain near zero throughout, and, importantly, do not display the progressive growth that would indicate unstable error accumulation under recursive inference. To verify that this accuracy is not specific to a single source position, Figure~\ref{fig3} repeats the comparison for a source at $x = 0.64$~km, following the same column arrangement. The predicted snapshots again match the reference closely and the difference panels remain negligible, confirming that the snapshot index and conditioning channels allow the operator to adapt to varying source locations without retraining.

Beyond individual snapshots, a more important product of simulation are the reliable surface records over the full recording time. Figure~\ref{fig4} shows the one second shot gathers for the two sources, with the FD reference in the left column, the recursively generated prediction in the middle column, and their difference in the right column. The predicted gathers capture the direct arrival, the subsequent refracted and reflected events, and their relative amplitudes. The residuals are weak and are concentrated near the direct arrival, where wavefield amplitudes and spatial gradients are largest. Later in the record the difference is essentially indistinguishable from zero.

For a more detailed view, Figure~\ref{fig5} extracts individual traces at $x = 0.4$~km and $x = 0.8$~km from both shot gathers. The predicted traces (dashed) overlie the FD references (solid) in both phase and amplitude across the full time window, for the near-offset and far-offset traces alike. Minor amplitude discrepancies are visible at isolated extrema, such as at the main trough of the far-offset trace for the $x = 0.32$~km source, but these are small relative to the event amplitudes and do not accumulate over time.

\begin{figure}[!htbp]
\centering
\includegraphics[width=0.5\textwidth]{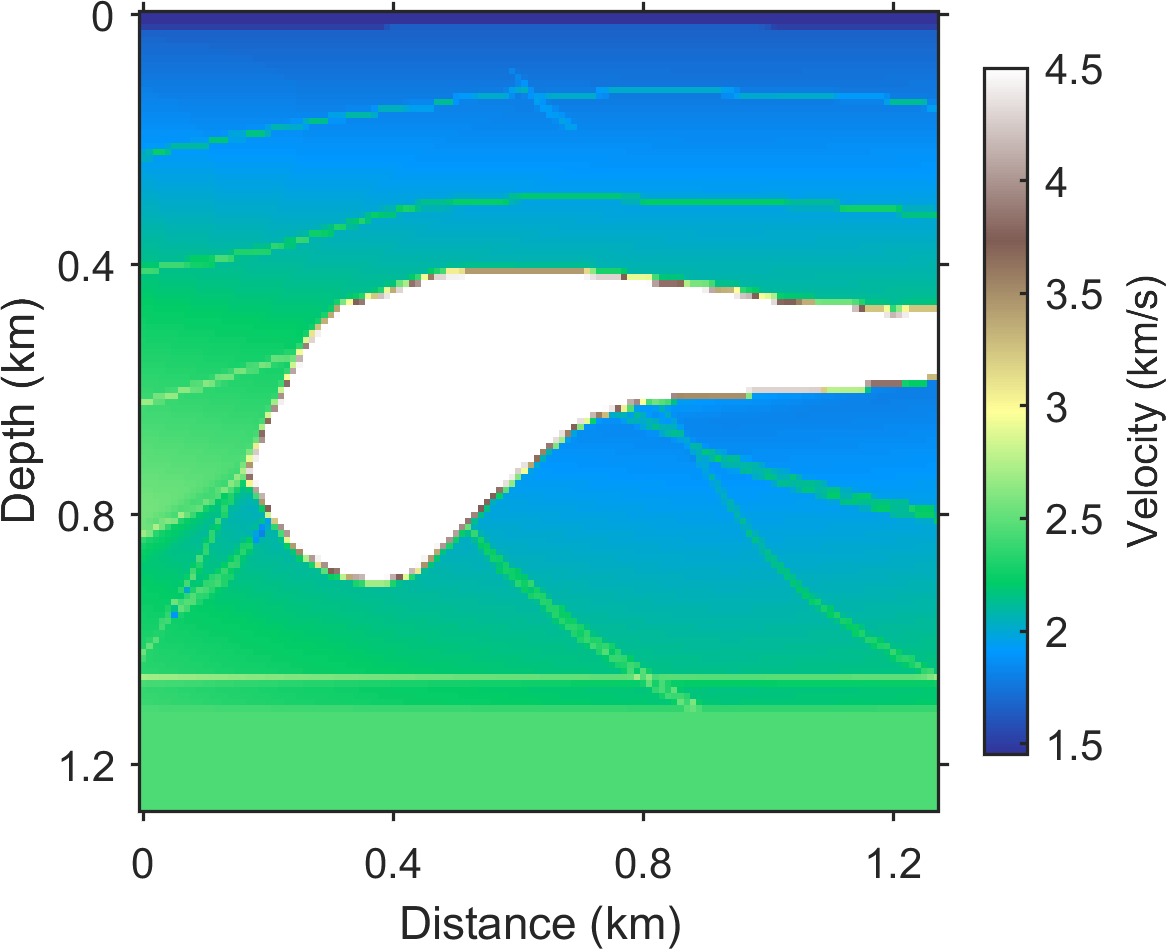}
\caption{The SEG/EAGE velocity model used for in-distribution testing, resized to $128 \times 128$ with a spatial sampling of $10$~m.}
\label{fig6}
\end{figure}

\begin{figure}[!htbp]
\centering
\includegraphics[width=1\textwidth]{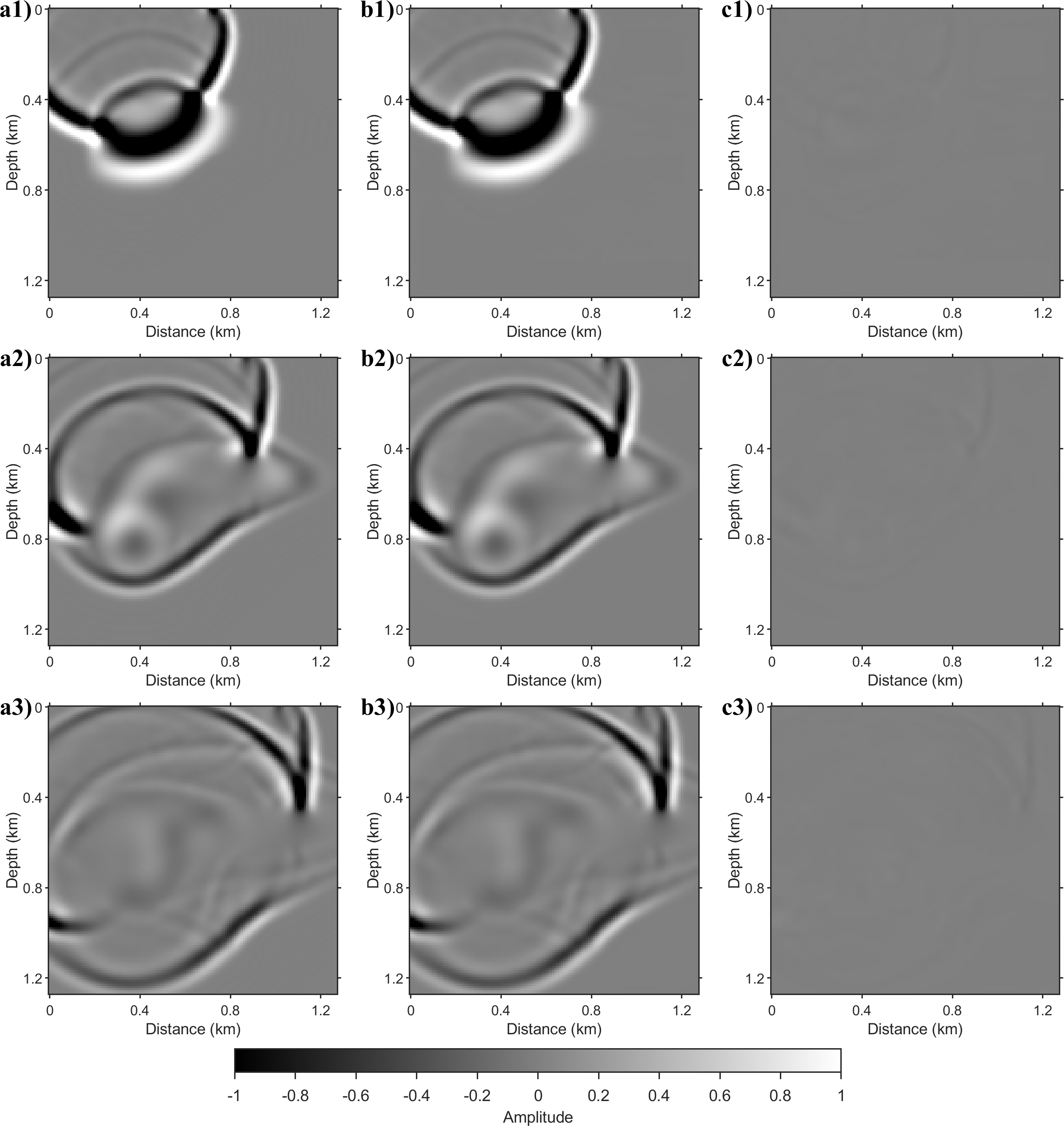}
\caption{Wavefield snapshot comparison for the SEG/EAGE model with a source at $x = 0.32$~km on the surface. Rows from top to bottom correspond to the wavefields at $0.3$, $0.4$, and $0.5$~s. (a1)-(a3) FD reference snapshots. (b1)-(b3) Snapshots generated recursively by the proposed method starting from $t = 0$ s. (c1)-(c3) Difference between the reference and the prediction.}
\label{fig7}
\end{figure}

\begin{figure}[!htbp]
\centering
\includegraphics[width=1\textwidth]{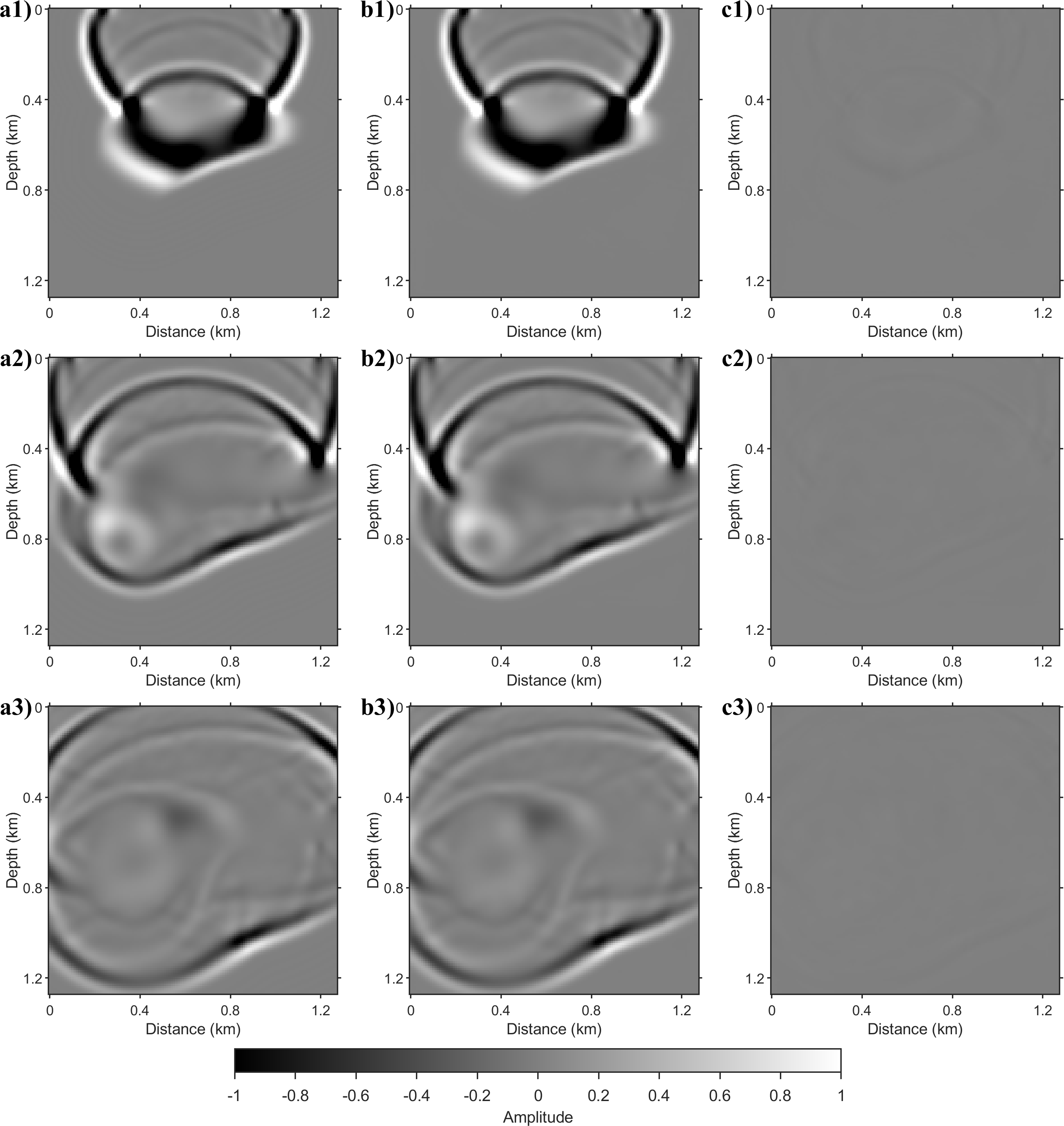}
\caption{Wavefield snapshot comparison for the SEG/EAGE model with a source at $x = 0.64$~km on the surface, following the same arrangement as Figure~\ref{fig7}.}
\label{fig8}
\end{figure}

\begin{figure}[!htbp]
\centering
\includegraphics[width=0.95\textwidth]{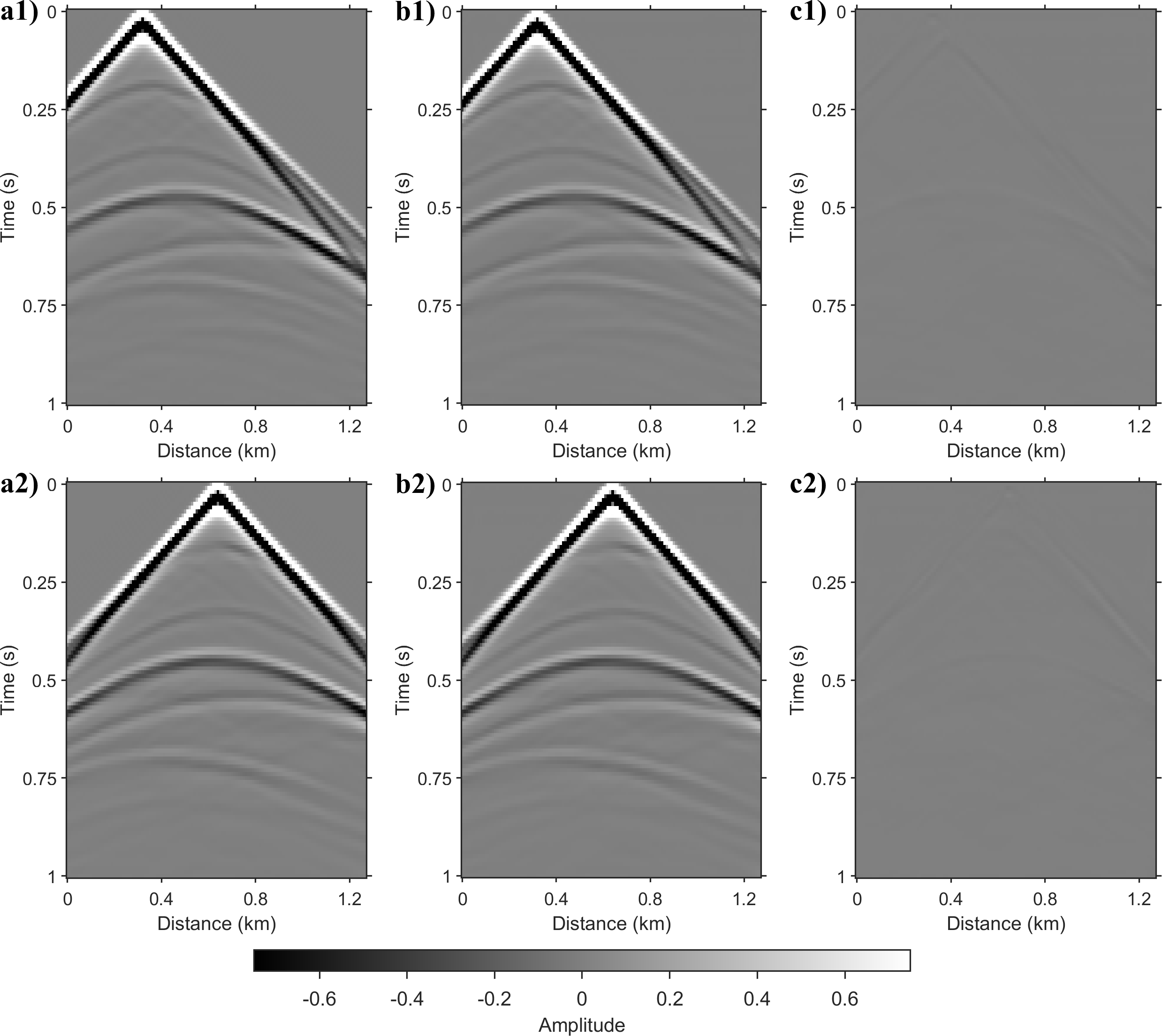}
\caption{Shot-gather comparison over $0$-$1$~s for the SEG/EAGE model. Rows correspond to the sources at $x = 0.32$~km (top) and $x = 0.64$~km (bottom). (a1)-(a2) FD reference gathers. (b1)-(b2) Gathers generated by the recursive inference of the proposed method. (c1)-(c2) Difference between the reference and the prediction.}
\label{fig9}
\end{figure}

\begin{figure}[!htbp]
\centering
\includegraphics[width=0.95\textwidth]{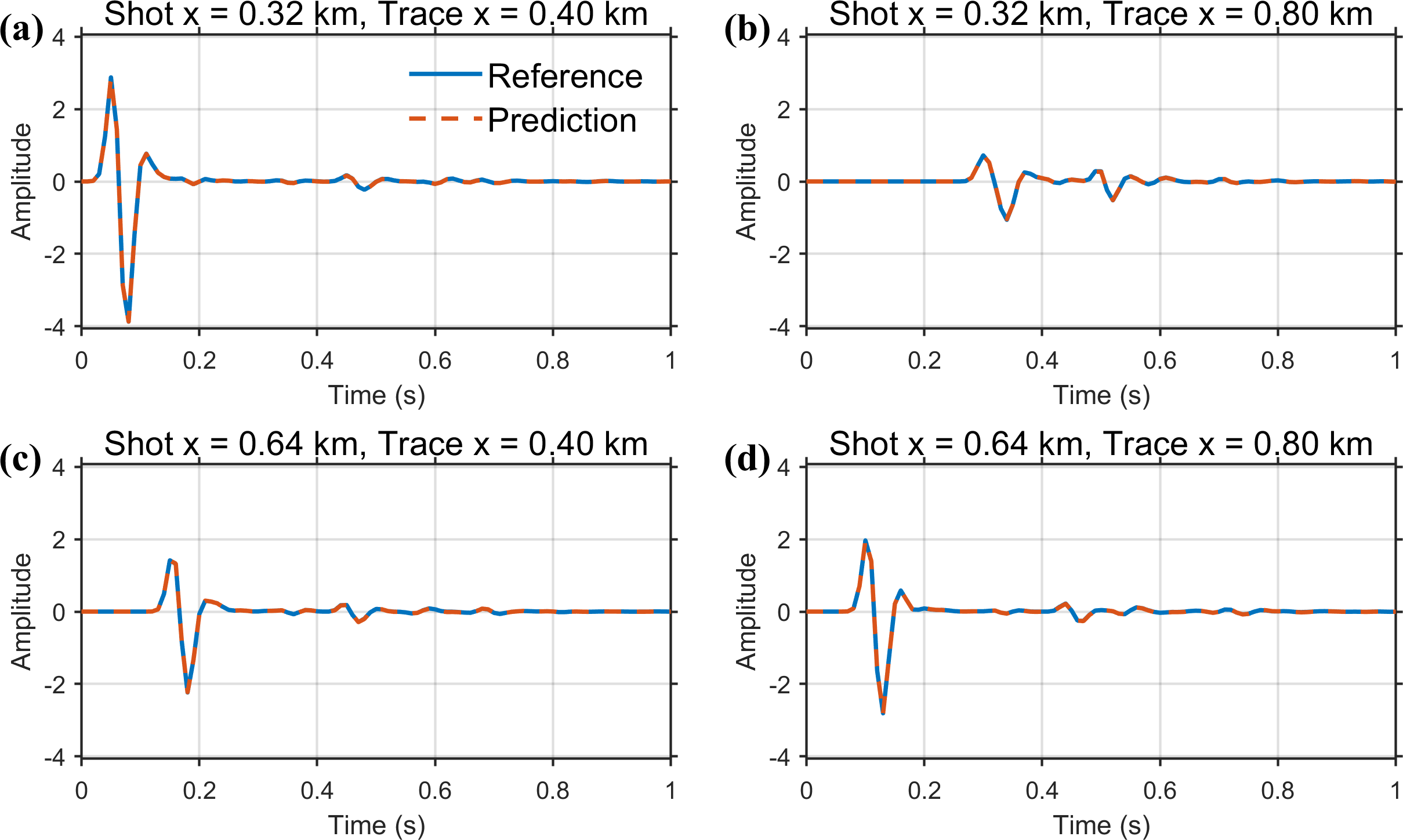}
\caption{Single-trace comparison for the SEG/EAGE model. Traces are extracted at $x = 0.4$~km and $x = 0.8$~km from the shot gathers of the two sources: the top row corresponds to the source at $x = 0.32$~km and the bottom row to the source at $x = 0.64$~km. Solid blue lines denote the FD reference and dashed orange lines denote the prediction of the proposed method.}
\label{fig10}
\end{figure}

\subsection{In-distribution test: the SEG/EAGE model}\label{sec:examples:sub3}

The second in-distribution test is the SEG/EAGE model. Figure~\ref{fig6} shows the SEG/EAGE velocity model used for testing, resized to $128 \times 128$ with a spatial sampling of $10$~m. The model is dominated by a high-velocity salt body of roughly $4.5$~km/s embedded within slower sedimentary layers. The strong impedance contrast at the salt boundary produces pronounced reflected and refracted phases, and the irregular shape of the body gives rise to complex scattering, making this a demanding test of the learned propagator.

Similar to the Overthrust model test, we first consider wavefield snapshots for a source at $x = 0.32$~km on the surface. Figure~\ref{fig7}, from top to bottom, compares the wavefield snapshots at $0.3$, $0.4$, and $0.5$~s, with the FD reference in the left column, the recursively generated prediction in the middle column, and their difference in the right column. The prediction closely matches the reference  at every time instant. The primary wavefront, the strong reflection off the salt boundary, and the scattered coda within and around the body are all recovered with correct position and amplitude. The difference panels remain near zero and show no progressive growth across the recursion, indicating that the high-contrast salt boundary does not destabilize the recursive inference. Figure~\ref{fig8} repeats the comparison for a source at $x = 0.64$~km, following the same column arrangement. The agreement between prediction and reference is again close, and the difference panels remain negligible, confirming that the operator adapts to varying source positions even in the presence of the salt body.

Figure~\ref{fig9} shows the one second long shot gathers for the two sources, with the FD reference, the recursively generated prediction, and their difference arranged by column. The predicted gathers capture the direct arrival together with the reflected and refracted events associated with the salt body, and their relative amplitudes are preserved. As in the Overthrust test, the residuals are weak and are concentrated near the direct arrival, while the later part of the record is essentially free of visible error.

Figure~\ref{fig10} provides a detailed comparison through individual traces extracted at $x = 0.4$~km and $x = 0.8$~km from both shot gathers. The predicted traces (dashed) closely follow the FD references (solid) in both phase and amplitude over the full recording time, for near-offset and far-offset traces alike. Only minor amplitude differences appear at isolated extrema, and these do not accumulate over the record.

Taken together, both in-distribution results demonstrate that, for geology consistent with the training distribution, the conditional diffusion model functions as an accurate forward-modeling operator. It reproduces both the propagating wavefield and the resulting surface records over long recursive horizons and across source positions.

\begin{figure}[!htbp]
\centering
\includegraphics[width=0.5\textwidth]{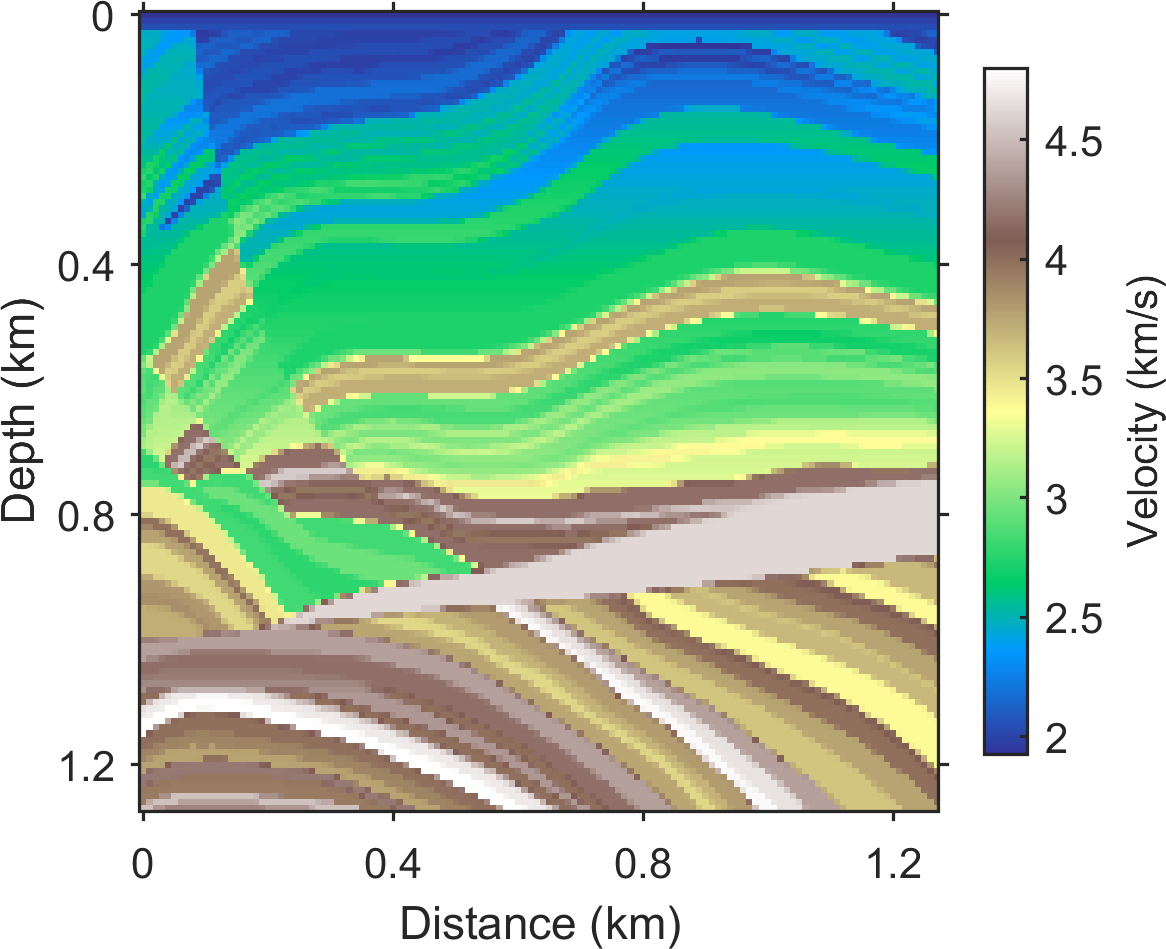}
\caption{The Marmousi velocity model used for out-of-distribution testing, resized to $128 \times 128$ with a spatial sampling of $10$~m.}
\label{fig11}
\end{figure}

\begin{figure}[!htbp]
\centering
\includegraphics[width=1\textwidth]{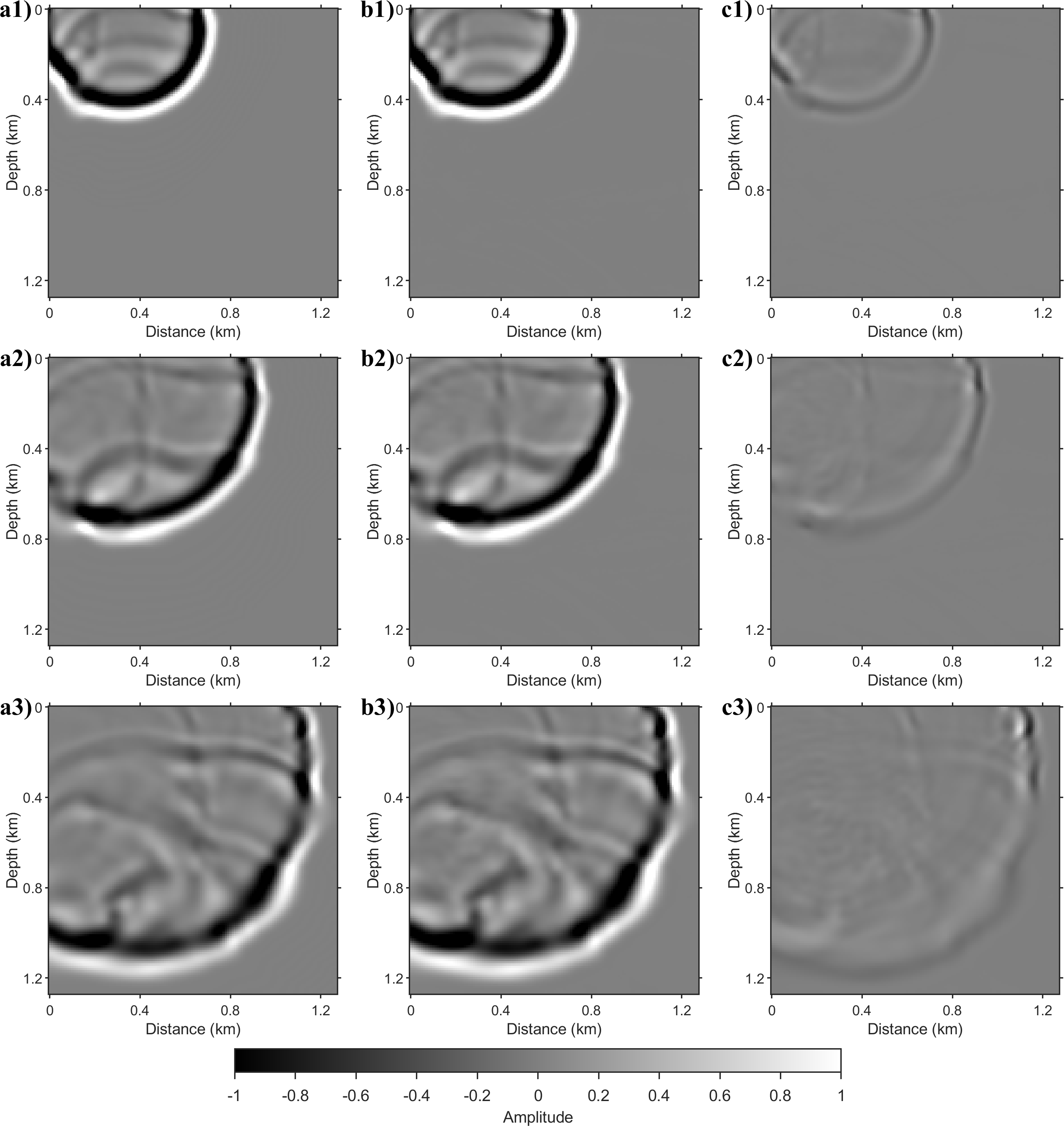}
\caption{Wavefield snapshot comparison for the Marmousi model with a source at $x = 0.32$~km on the surface. Rows from top to bottom correspond to the wavefields at $0.2$, $0.3$, and $0.4$~s. (a1)-(a3) FD reference snapshots. (b1)-(b3) Snapshots generated recursively by the proposed method starting from $t = 0$ s. (c1)-(c3) Difference between the reference and the prediction.}
\label{fig12}
\end{figure}

\begin{figure}[!htbp]
\centering
\includegraphics[width=1\textwidth]{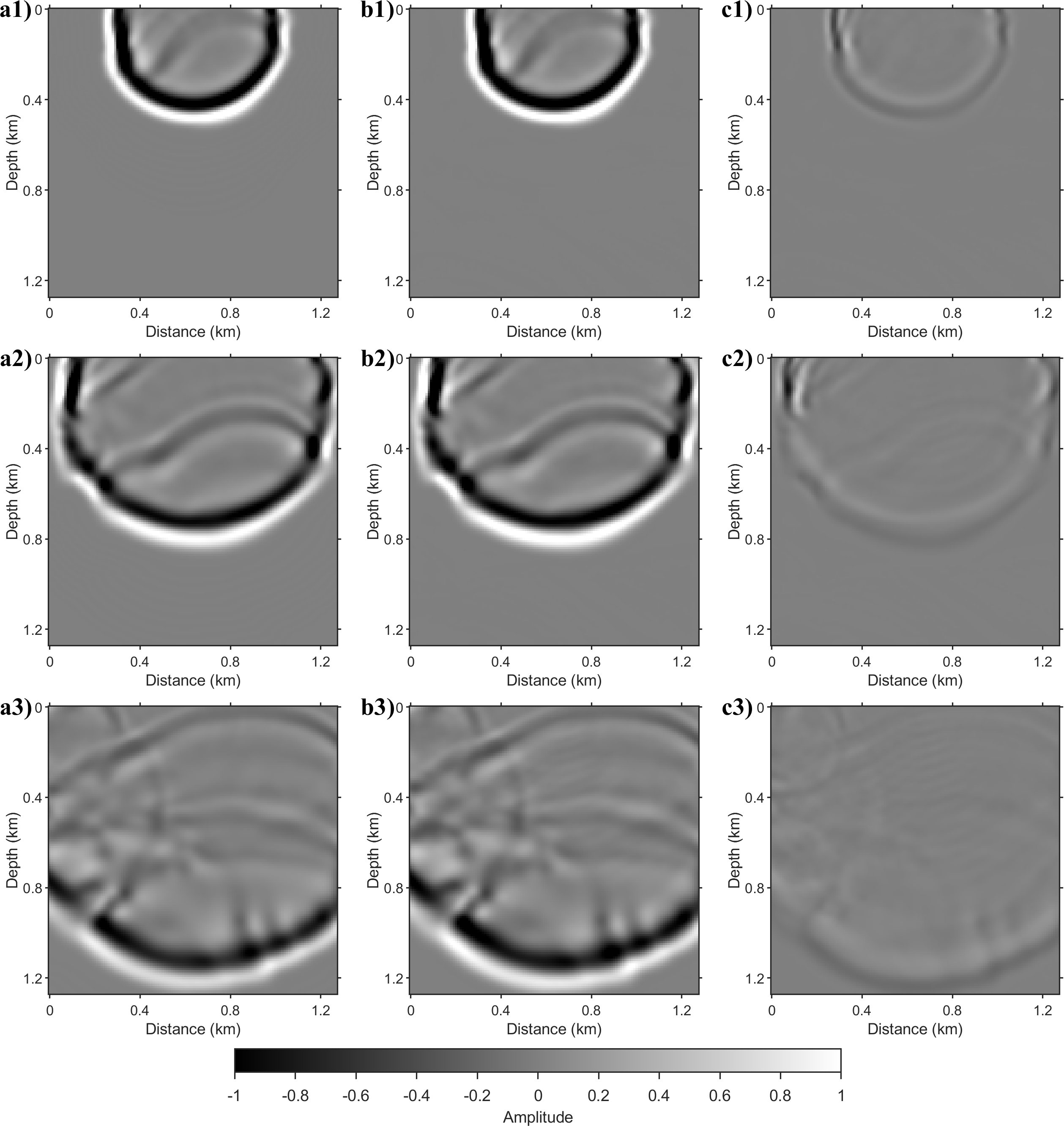}
\caption{Wavefield snapshot comparison for the SEG/EAGE model with a source at $x = 0.64$~km on the surface, following the same arrangement as Figure~\ref{fig12}. }
\label{fig13}
\end{figure}

\begin{figure}[!htbp]
\centering
\includegraphics[width=0.95\textwidth]{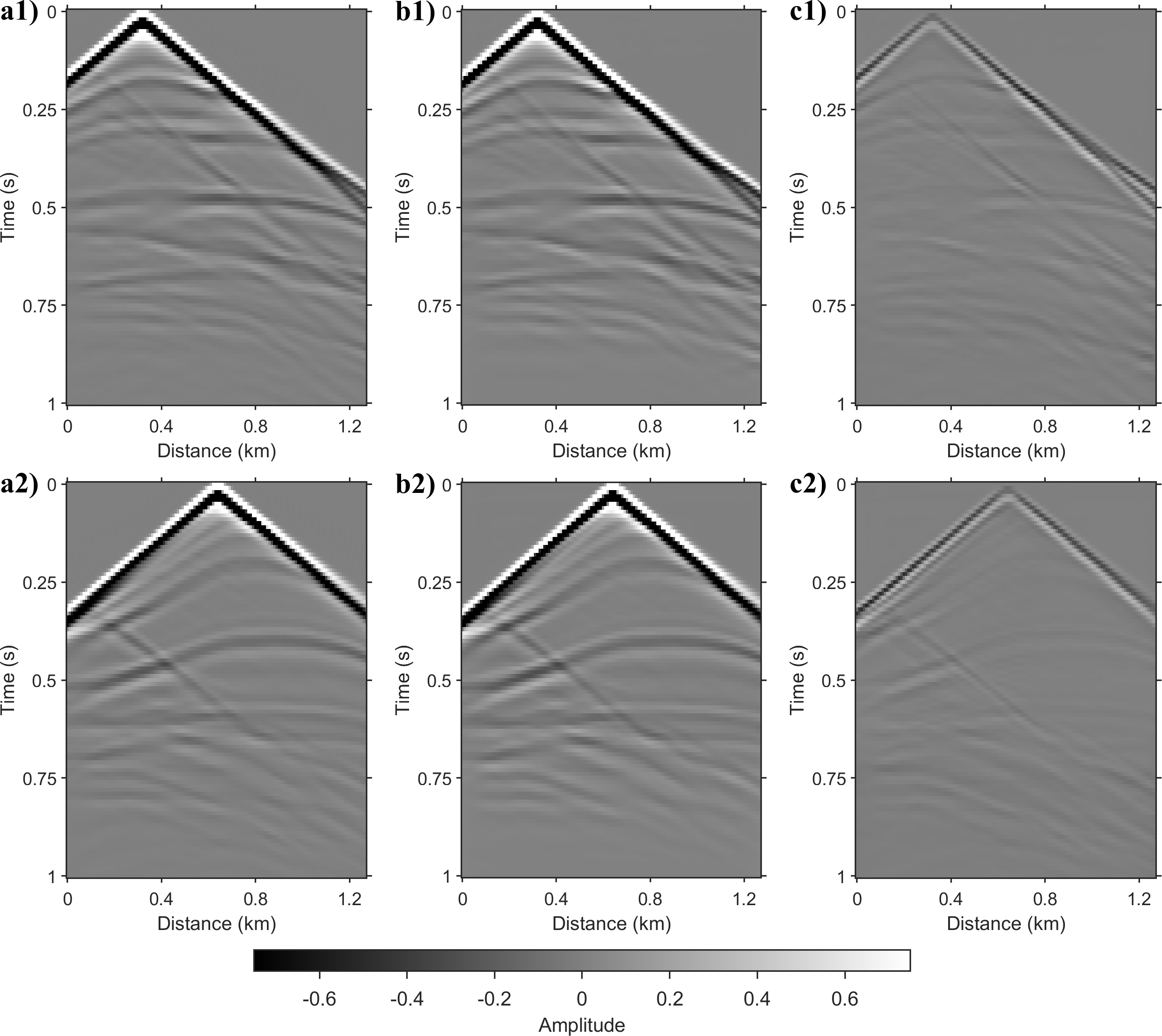}
\caption{Shot-gather comparison over $0$-$1$~s for the Marmousi model. Rows correspond to the sources at $x = 0.32$~km (top) and $x = 0.64$~km (bottom). (a1)-(a2) FD reference gathers. (b1)-(b2) Gathers generated by the recursive inference of the proposed method. (c1)-(c2) Difference between the reference and the prediction.}
\label{fig14}
\end{figure}

\begin{figure}[!htbp]
\centering
\includegraphics[width=0.95\textwidth]{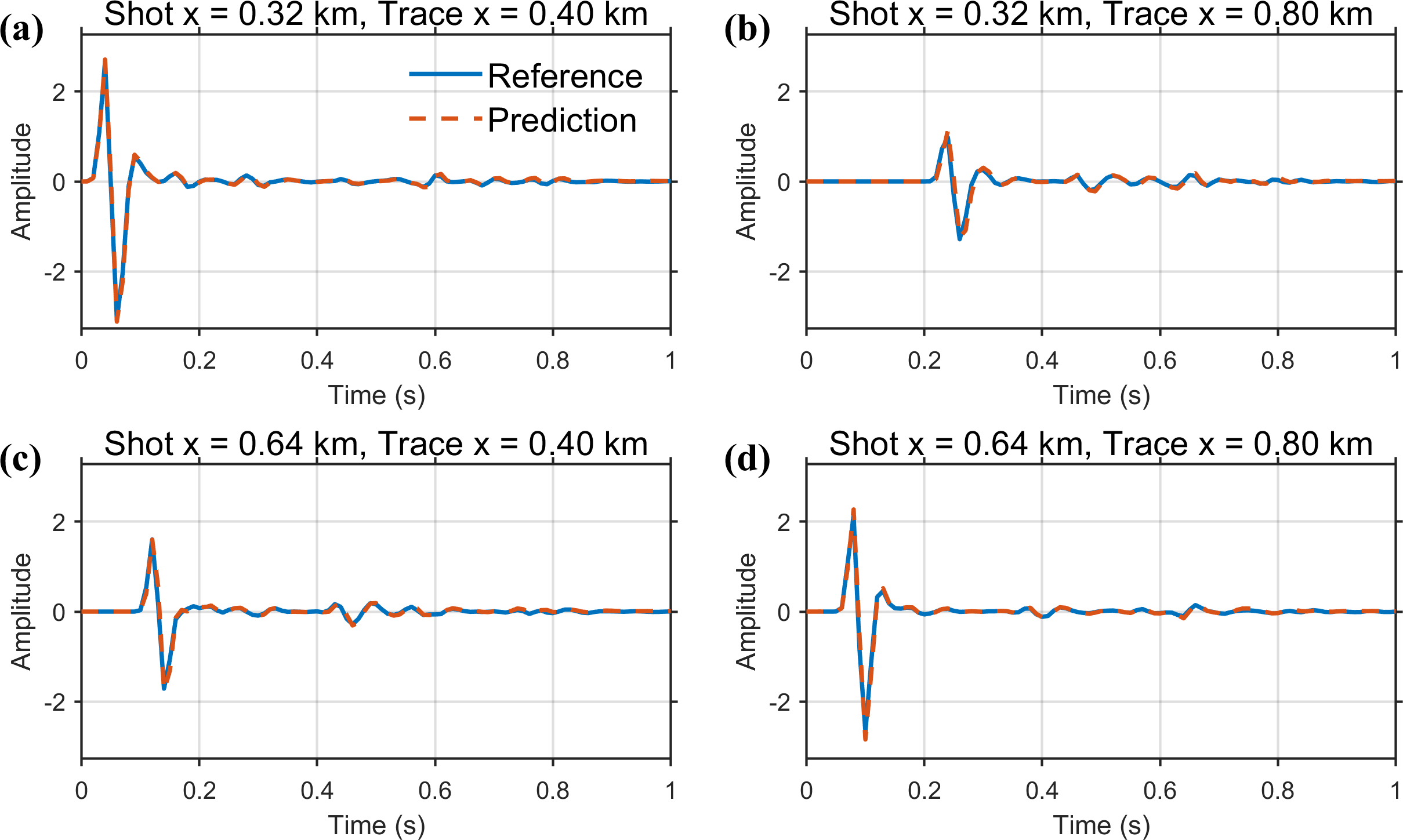}
\caption{Single-trace comparison for the Marmousi model. Traces are extracted at $x = 0.4$~km and $x = 0.8$~km from the shot gathers of the two sources: the top row corresponds to the source at $x = 0.32$~km and the bottom row to the source at $x = 0.64$~km. Solid blue lines denote the FD reference and dashed orange lines denote the prediction of the proposed method.}
\label{fig15}
\end{figure}

\subsection{Out-of-distribution test: the Marmousi model}\label{sec:examples:sub4}

The two preceding tests probe geology that is represented, in patch form, within the training set. A more strict assessment is whether the learned operator generalizes to a velocity model whose structural style is absent from training. For this purpose we use the Marmousi model, extracting a portion from its right side and resizing it to $128 \times 128$ to match the input dimension of the network. Importantly, the Marmousi model does not contribute to the training set, so this experiment constitutes a reasonable out-of-distribution test, and we examine its results with attention to the effects of this distribution shift. Figure~\ref{fig11} shows the Marmousi velocity model used for testing, which is characterized by numerous thin, steeply dipping layers and a complex faulted structure.

Again, we first examine wavefield snapshots for a source at $x = 0.32$~km on the surface. Figure~\ref{fig12} compares the wavefields at $0.2$, $0.3$, and $0.4$~s, with the FD reference, the recursive prediction, and their difference arranged by column. The prediction reproduces the principal features of the reference, where the position and curvature of the main wavefront, the strong early reflections, and the overall pattern of the scattered coda are all broadly recovered. Compared with the in-distribution tests, however, the difference panels are no longer close to zero. Weak residual energy is visible along the main wavefront and within the coda, and it becomes somewhat more pronounced for the horizontally traveling parts of the wavefield. The residual nonetheless remains small relative to the wavefield amplitude, and the predicted wavefront does not drift in position or develop spurious phases.

Figure~\ref{fig13} shows the corresponding comparison for a source at $x = 0.64$~km on the surface. The behavior is consistent with the first source: the predicted snapshots track the reference wavefronts and reflections, while the difference panels again show a weak but non-negligible residual that is most visible near the strongest events. The fact that the error pattern is similar for both source positions indicates that the distribution shift, rather than any particular source geometry, is the dominant factor.

Figure~\ref{fig14} presents the one second long shot gathers for the two sources. The predicted gathers capture the direct arrival and the main sequence of reflected and refracted events, with their traveltimes and overall amplitude pattern preserved. Consistent with the snapshot comparisons, the difference panels are more visible than in the in-distribution cases: the residual is concentrated along the direct arrival and the more energetic events, where the finely layered structure produces rapid spatial variation that the operator reproduces slightly less accurately. The later, weaker part of the record remains close to the reference.

Figure~\ref{fig15} provides the single-trace comparison at $x = 0.4$~km and $x = 0.8$~km for both shot gathers. The predicted traces follow the FD references closely in traveltime and phase across the full recording window, and the amplitudes of the major events are well matched. Small amplitude discrepancies are visible at certain extrema, slightly larger than those observed in the in-distribution tests, but the traces do not show phase drift or growing misfit over time.

These results show the expected signature of a distribution shift: relative to the in-distribution tests, the prediction error on Marmousi is visibly larger and accumulates mildly along the recursion. Crucially, however, the degradation is gradual rather than catastrophic. The learned operator continues to reproduce wavefront geometry, event traveltimes, and phase behavior reliably, and the recursive inference remains stable over the full simulation horizon. This indicates that the conditional diffusion model has learned a propagation operator with a meaningful degree of generality, retaining usable forward-modeling accuracy even on geology well outside its training distribution.

\begin{figure}[!htbp]
\centering
\includegraphics[width=0.6\textwidth]{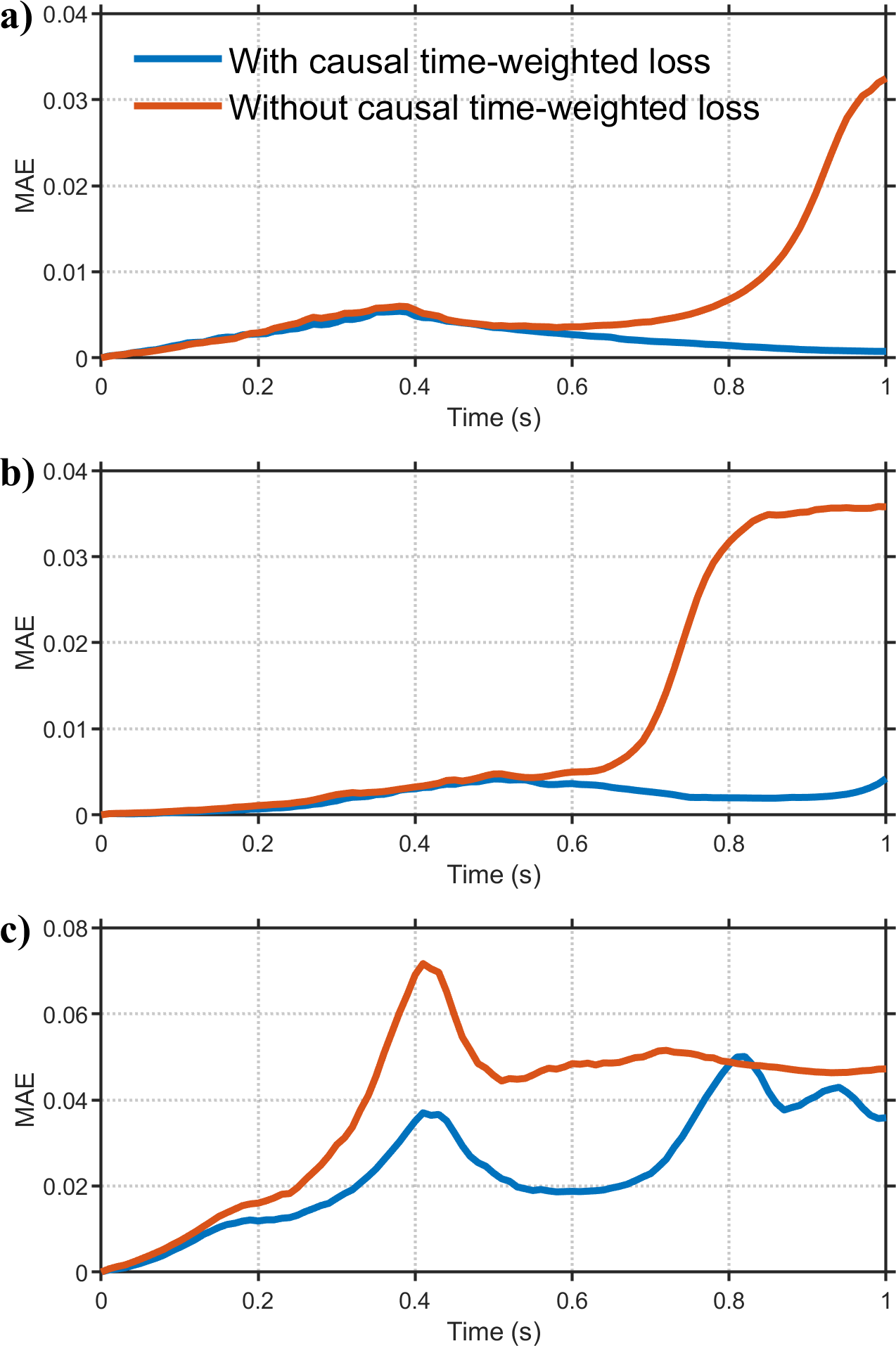}
\caption{Per-snapshot MAE as a function of physical time, for the source at $x = 0.32$~km, comparing training with (blue) and without (orange) the causal time-weighted loss. (a) Overthrust model. (b) SEG/EAGE model. (c) Marmousi model.}
\label{fig16}
\end{figure}

\subsection{Effect of the causal time-weighted loss}\label{sec:examples:sub5}

The preceding tests demonstrate the overall accuracy and generalization behavior of the proposed model, but they do not isolate the contribution of the causal time-weighted loss introduced in Section~\ref{sec:method:sub3:sub3}. To assess this design choice directly, we retrain the network with an identical configuration, i.e., the same architecture, dataset, optimizer, learning rate, batch size, and number of iterations as in Section~\ref{sec:examples:sub1}, with only the loss function modified, reverting to the uniformly weighted MSE objective of equation~\ref{eq:base_loss}. All subsequent comparisons hold every other factor fixed, so that any difference in performance is attributable to the weighting scheme alone.

We quantify accuracy with two complementary metrics computed over the $100$ predicted wavefield snapshots covering $0$-$1$~s for each source. Let $u^{n}_{\text{ref}}$ denote the reference snapshot from the FD solver and $\hat{u}^{n}$ the corresponding recursive prediction. The mean absolute error (MAE) is defined as
\begin{equation}\label{eq:mae}
\text{MAE} \;=\; \frac{1}{N \, |\Omega|} \sum_{n=1}^{N} \sum_{\mathbf{x} \in \Omega} \bigl|\, u^{n}_{\text{ref}}(\mathbf{x}) - \hat{u}^{n}(\mathbf{x}) \,\bigr|,
\end{equation}
where $\Omega$ is the spatial domain, $|\Omega|$ is the number of grid points, and $N = 100$. The signal-to-noise ratio (SNR), reported in decibels, is defined as
\begin{equation}\label{eq:snr}
\text{SNR} \;=\; 10 \, \log_{10} \!\left( \frac{\sum_{n,\mathbf{x}} \bigl|\, u^{n}_{\text{ref}}(\mathbf{x}) \,\bigr|^{2}}{\sum_{n,\mathbf{x}} \bigl|\, u^{n}_{\text{ref}}(\mathbf{x}) - \hat{u}^{n}(\mathbf{x}) \,\bigr|^{2}} \right).
\end{equation}
Table~\ref{tab1} reports both metrics for all three test models and both source positions, with and without the causal time-weighted loss.

\begin{table}[!htbp]
\centering
\caption{Wavefield prediction accuracy with and without the causal time-weighted loss, evaluated on the $100$ predicted snapshots covering $0$--$1$~s. ``Without'' denotes the baseline trained with the uniformly weighted MSE objective; ``With'' denotes the proposed causal time-weighted objective. All other training factors are held fixed. Lower MAE and higher SNR indicate better accuracy.}
\label{tab1}
\begin{tabular}{llcccc}
\toprule
\multirow{2}{*}{Model} & \multirow{2}{*}{Source} & \multicolumn{2}{c}{MAE} & \multicolumn{2}{c}{SNR (dB)} \\
\cmidrule(lr){3-4} \cmidrule(lr){5-6}
 & & Without & With & Without & With \\
\midrule
\multirow{2}{*}{Overthrust} & $x = 0.32$~km & $0.0068$ & $\mathbf{0.0024}$ & $18.59$ & $\mathbf{30.81}$ \\
                            & $x = 0.64$~km & $0.0063$ & $\mathbf{0.0020}$ & $20.39$ & $\mathbf{33.58}$ \\
\midrule
\multirow{2}{*}{SEG/EAGE}   & $x = 0.32$~km & $0.0113$ & $\mathbf{0.0022}$ & $13.29$ & $\mathbf{33.70}$ \\
                            & $x = 0.64$~km & $0.0111$ & $\mathbf{0.0029}$ & $13.74$ & $\mathbf{25.69}$ \\
\midrule
\multirow{2}{*}{Marmousi}   & $x = 0.32$~km & $0.0381$ & $\mathbf{0.0239}$ & $8.09$  & $\mathbf{10.81}$ \\
                            & $x = 0.64$~km & $0.0373$ & $\mathbf{0.0245}$ & $7.98$  & $\mathbf{10.53}$ \\
\bottomrule
\end{tabular}
\end{table}

Several observations follow from Table~\ref{tab1}. First, the causal time-weighted loss improves accuracy in every single test configuration, on both in-distribution and out-of-distribution models and at both source positions, without exception. Second, the improvement on in-distribution geology is substantial: on Overthrust, the MAE is reduced by roughly a factor of three and the SNR increased by $12$--$13$~dB, while on SEG/EAGE the MAE is reduced by roughly a factor of five and the SNR improved by as much as $20$~dB. These gains correspond to a qualitative change in long-horizon accuracy rather than a marginal refinement. Second, the table also quantifies the effect of distribution shift quantitatively explicit. With the causal time-weighted loss in place, the MAE on the in-distribution Overthrust and SEG/EAGE models is around $0.002$, whereas on the out-of-distribution Marmousi model it rises to roughly $0.024$. Correspondingly, the SNR drops from above $25$~dB on in-distribution geology to around $10$-$11$~dB on Marmousi. This pattern is consistent with the visual observations of Section~\ref{sec:examples:sub4}: the operator continues to track wavefronts and main events on Marmousi, but with a measurably higher residual that reflects the gap between training and test distributions. Third, even in this out-of-distribution regime, the causal time-weighted loss continues to deliver a clear improvement: it reduces the MAE on Marmousi by about $37\%$ and improves the SNR by roughly $2.5$~dB for both sources. The benefit of the weighting scheme is therefore not specific to in-distribution geology; it is a property of the training procedure that carries over to unseen velocity models.

To understand the mechanism behind these aggregate improvements, we examine how the prediction error evolves along the recursion. Figure~\ref{fig16} shows, for the source at $x = 0.32$~km, the snapshot-wise MAE plotted as a function of physical time accross the three test models, comparing training with and without the causal weighting. On the two in-distribution models (panels~a and~b), the difference between the two training strategies is striking. Without the causal weighting, the error grows slowly at first and then accelerates sharply in the later part of the recursion, rising from below $0.005$ early in the simulation to around $0.03$--$0.036$ near $1$~s. With the causal weighting, the error remains essentially flat across the full simulation horizon, reaching its maximum near the middle of the time window and then decaying back toward small values. This is exactly the behavior predicted by the design of the loss in Section~\ref{sec:method:sub3:sub3}: by penalizing early-snapshot errors more strongly during training, the weighting suppresses the source of the recursive error amplification, and the long-horizon accuracy follows as a consequence. On the out-of-distribution Marmousi model (panel~c), the picture is more nuanced. Both curves now exhibit noticeable fluctuations, including a pronounced peak near $0.4$~s, which we attribute to the distribution shift rather than to either training strategy. Even so, the curve obtained with the causal time-weighted loss lies below the baseline curve over nearly the entire time window, and its peak value is much lower than the baseline peak. Therefore, the causal weighting still provides a consistent improvement throughout the recursive prediction process, even when the underlying distribution-shift error cannot be fully removed by training-time strategies alone.

As a result, the table and the per-time error curves confirm that the causal time-weighted loss is not merely a simple adjustment, but a key component of the proposed framework. It directly reduces the error accumulation caused by recursive inference and leads to substantial improvements in long-term prediction accuracy for the geological settings included in training, while still providing benefits when the operator is asked to make predictions beyond its training distribution.
\section{\textbf{Discussion}}\label{sec:discussion}
In this section, we first analyze the property underlying the reported speedup: the decoupling of the inference time step from the CFL stability condition, together with the corresponding multi-shot scaling behavior. We then examine the mechanism through which the causal time-weighted loss delivers its observed benefits, framing it in terms of one-step error propagation along the recursion. Finally, we examine what the contrast between in-distribution and out-of-distribution performance tells us about the generalization capacity of the learned propagator, and which avenues are most promising for closing the remaining gap.

\subsection{Decoupling of training and inference time steps}\label{sec:discussion:sub1}

The most consequential property of the learned propagator $\mathcal{P}_{\Delta t}$ defined in equation~\ref{eq:operator} is that the physical time increment at which it advances the wavefield is no longer governed by the stability condition of equation~\ref{eq:cfl}. The training data are generated by a finite-difference (FD) solver with time step $\Delta t_{\text{FD}} = 0.001$~s, chosen to satisfy both the dispersion condition (equation~\ref{eq:dispersion}) and the CFL condition (equation~\ref{eq:cfl}). These constraints are unavoidable for the offline data-generation stage. Adjacent training pairs $(\mathbf{u}^{n-4:n}, u^{n+1})$, however, are not formed from consecutive FD snapshots. We extract every tenth snapshot from the FD time series, so that the model learns a propagator whose effective physical time increment is
\begin{equation}\label{eq:dt_scaling}
\Delta t \;=\; 10 \,\Delta t_{\text{FD}} \;=\; 0.01~\text{s}.
\end{equation}
The CFL condition does not constrain $\Delta t$ at inference time. The bound on the achievable $\Delta t$ is instead determined by data sufficiency, in the sense that the training pairs must retain sufficient predictive correlation --- $u^{n+1}$ must remain predictable from the recent snapshot history $\mathbf{u}^{n-4:n}$ alone. As $\Delta t$ grows, the wavefield change between consecutive training snapshots becomes larger and the corresponding mapping becomes increasingly nonlinear, so the choice of $\Delta t$ trades acceleration against the capacity demanded of $f_\theta$. 

The resulting computational comparison with the conventional FD method is direct. Propagating the wavefield over $0$-$1$~s in the $128 \times 128$ test setting of Section~\ref{sec:examples} requires $1000$ time-stepping iterations under the tenth-order staggered-grid solver and $100$ recursive forward passes under our method. To make the comparison concrete, we benchmarked both procedures end-to-end on a single NVIDIA A100 GPU, with all initialization, PML setup (a $50$-grid-point absorbing layer surrounding the $128 \times 128$ domain), boundary updates, and snapshot storage included in the timing. Under matched hardware and matched physical configuration, the FD solver completes a single-shot simulation in $5.87$~s, while the proposed method completes the same simulation in $2.71$~s, an end-to-end speedup of approximately $2.17\times$. We emphasize that this comparison is conservative in two respects. First, the FD baseline is itself GPU-accelerated through a PyTorch convolution-based implementation of the staggered-grid stencil, rather than a CPU reference. Against a CPU FD implementation, the relative speedup would be substantially larger. Second, the $\Delta t = 10\,\Delta t_{\text{FD}}$ used here is a deliberately moderate choice. The underlying framework admits larger $\Delta t$ at the cost of increased nonlinearity of the learned mapping, which would translate directly into a higher step-count ratio.

A practical advantage that the wall-clock comparison above does not capture is the behavior of the two methods under multi-shot workloads, which dominate the total compute budget of iterative inversion pipelines such as full waveform inversion and reverse-time migration. The proposed method handles independent source positions as a batch dimension of the neural network, and the resulting per-shot cost grows far more slowly than linearly with batch size, where doubling the number of shots increases the inference time by substantially less than a factor of two, thanks to the inherent batch parallelism of the GPU. An FD solver, by contrast, scales essentially linearly: each additional shot requires its own copy of the PML state, source coupling, and time-stepping logic, and the per-shot cost remains roughly constant. The relative advantage of the proposed framework therefore widens as the number of shots per inversion iteration grows, which is precisely the regime of practical interest.

We close this subsection by clarifying the scope of what the present comparison establishes. The speedup we report is achieved by replacing the explicit time-stepping operator $F$ of equation~\ref{eq:fdm} with the learned operator $\mathcal{P}_{\Delta t}$ of equation~\ref{eq:operator}, and corresponds specifically to a relaxation of the temporal stability constraint of equation~\ref{eq:cfl}. An analogous relaxation along the spatial axis is, however, equally accessible within the same framework: an FD solver must refine its spatial grid to control numerical dispersion (equation~\ref{eq:dispersion}), but the training data for our method can be generated on a fine FD grid and then downsampled to a coarser grid on which the network is trained, so that the learned propagator operates at a spatial sampling that would itself be dispersion-limited under an FD scheme. The acceleration achievable by our framework is therefore not confined to the time axis. Rather, the analysis we report here is one instance of a more general strategy in which the cost of stability and accuracy is transferred from inference to a one-time training phase.

\subsection{Why causal time-weighting matters: a mechanistic view}\label{sec:discussion:sub2}

The ablation study of Section~\ref{sec:examples:sub5} established empirically that the causal time-weighted loss substantially improves accuracy and prevents the late-time error blow-up visible in the baseline (Figure~\ref{fig16}). We now examine why this weighting scheme is particularly well suited to the wavefield recursion problem.

The root cause of the asymmetry between snapshot indices is the recursive nature of inference. Let $\varepsilon_n = \hat{u}^{n} - u^{n}_{\text{ref}}$ denote the prediction error at snapshot $n$. At the next step, the network consumes $\hat{u}^{n}$ rather than $u^{n}_{\text{ref}}$ as its conditioning input, so to first order
\begin{equation}\label{eq:error_propagation}
\varepsilon_{n+1} \;\approx\; \tilde{\varepsilon}_{n+1} \;+\; J_n\, \varepsilon_n,
\end{equation}
where $\tilde{\varepsilon}_{n+1}$ is the fresh one-step error at snapshot $n+1$ and $J_n$ is the local sensitivity of the propagator $f_\theta$ to its most recent conditioning snapshot. Although $\|J_n\|$ is typically close to unity for a propagator advancing a wavefield over a short interval, it can locally exceed unity at sharp wavefronts and high-contrast boundaries, so early-step errors can be substantially amplified by the time the recursion reaches its end. This is the mechanism behind the late-time error blow-up of the baseline curves in Figure~\ref{fig16}.

The causal time-weighted loss addresses this mechanism directly. By assigning a weight $\omega(n)$ that shrinks with the cumulative residual error at all snapshots strictly preceding $n$, the loss enforces an implicit training order in which snapshots near the source are learned first and later snapshots enter the active training set only once their predecessors have been fitted to acceptable accuracy. Because the early entries of $\varepsilon_n$ are precisely what feeds the propagation term $J_n\,\varepsilon_n$ in equation~\ref{eq:error_propagation}, strongly suppressing them has a multiplicative effect on long-horizon accuracy: a fixed reduction in $\|\varepsilon_n\|$ at small $n$ yields a far larger reduction in $\|\varepsilon_N\|$ at the end of the recursion than the same reduction applied at large $n$. The training trajectory thus proceeds along the same direction as physical wave propagation, accumulating competence outward in time from the source.

The weighting strategy we adopt is inspired by the causal training strategy for physics-informed neural networks (PINNs) of \cite{wang2022respecting}, but differs in one important respect. In the PINN setting, the temporal weights are computed from the current-iteration PDE residual and therefore reflect only the loss at the present training step. In our setting, the per-snapshot errors $L_{\text{ema}}[n]$ that drive the weights are exponential moving averages accumulated across training iterations, with only those entries corresponding to snapshots in the current minibatch being updated. The weights are therefore insensitive to the minibatch stochasticity in $(n, t)$ sampling, evolve smoothly, and aggregate information over the entire training history rather than fluctuating with the noise of any individual minibatch. Such behavior could not be reliably provided by a per-iteration adaptive scheme without temporal smoothing. From this perspective, the causal time-weighted loss is a training-time strategy specifically tailored to the structure of the wavefield recursion problem. It does not modify the model architecture, the inference procedure, or the underlying physics. Instead, it modifies only how training compute is allocated across snapshot indices, in a way that mirrors the causal direction of physical wave propagation itself.

\subsection{Generalization and the limits of in-distribution training}\label{sec:discussion:sub3}

The experiments of Sections~\ref{sec:examples:sub2}--\ref{sec:examples:sub4} reveal two distinct regimes. On Overthrust and SEG/EAGE, the operator functions as an accurate forward-modeling tool, with per-snapshot MAE remaining below $0.005$ across the full $0$--$1$~s recursion. On Marmousi, which lies outside the training distribution, the same operator continues to reproduce wavefront geometry, traveltimes, and the main reflected and refracted phases, but with an order-of-magnitude larger MAE and an of SNR around $10$~dB.

Two features of this contrast are worth emphasizing. First, the degradation is graceful rather than catastrophic: predicted wavefronts do not drift in position, the recursion remains stable over the full horizon, and the causal time-weighted loss continues to provide a measurable improvement (Table~\ref{tab1}). This suggests that the model has not merely memorized propagation patterns specific to the training velocities, but has acquired a propagator with a meaningful degree of physical generality. Second, the prediction residual on Marmousi remains non-negligible for applications demanding high-fidelity wavefields, and this gap is the generic behavior of data-driven surrogates under distribution shift rather than a failure mode unique to our formulation.

Several routes are available to close this gap, all compatible with the framework presented here. Broadening the training distribution to include finely layered and steeply dipping structures would shift such media into the in-distribution regime. Fine-tuning on a small number of training pairs from the target geology, generated on the fly by a finite-difference solver, can adapt the operator to a specific survey context at a fraction of the original training cost. Incorporating the wave equation as a soft constraint during training would bias the learned operator toward physically admissible solutions in regions of input space sparsely covered by the training data. The architecture and dataset used in this work were chosen to demonstrate the framework rather than to push its generalization frontier, and we therefore regard the Marmousi result as a baseline indication of the behavior under unfavorable conditions, not a definitive statement of the framework's capabilities.

\section{\textbf{Conclusions}}
We presented a conditional diffusion model for seismic wavefield simulation that formulates wavefield propagation as a recursive, physically conditioned state-transition process. The network predicts the next wavefield snapshot from a short history of recent wavefield snapshots, the velocity model, and the snapshot index, and the network is trained with a causal time-weighted loss that aligns the training trajectory with the physical recursion and reduces long-horizon error accumulation. During inference, the network directly generates the clean next wavefield snapshot in a single forward pass, avoiding the costly iterative sampling procedure commonly associated with diffusion models. Numerical experiments on the Overthrust, SEG/EAGE, and Marmousi models demonstrate that the proposed framework delivers accurate forward modeling for in-distribution geology and remains stable under distribution shift. In particular, the Marmousi test shows that the predicted wavefronts remain physically coherent, although the residual errors become non-negligible compared with the in-distribution cases. This result suggests that the learned propagator captures a meaningful degree of transferable wave-propagation behavior, while also indicating the need for broader training distributions and improved generalization strategies. From a computational perspective, the framework advances the wavefield at a physical step ten times larger than that of the underlying finite-difference solver and achieves an end-to-end speedup of $2.17\times$ under matched hardware conditions. Future work will focus on reducing the in-distribution/out-of-distribution accuracy gap, incorporating more diverse velocity models and source-frequency conditions, and extending the framework to elastic and three-dimensional wave propagation.

\section*{Acknowledgments}
This publication is based on work supported by the King Abdullah University of Science and Technology (KAUST). The authors thank the DeepWave sponsors for their support. This work utilized the resources of the Supercomputing Laboratory at King Abdullah University of Science and Technology (KAUST) in Thuwal, Saudi Arabia.
\section*{Code and data availability}
The data and accompanying codes that support the findings of this study are available at: \url{https://github.com/DeepWave-KAUST/GenWP-pub}.

\bibliographystyle{unsrtnat}
\bibliography{references}

\end{document}